\documentclass[aip, cha, amsmath, amssymb, reprint, numerical]{revtex4-1}

\DeclareUnicodeCharacter{00F7}{$\div$}
\DeclareUnicodeCharacter{2209}{$\notin$}

\usepackage[frozencache=true, cachedir=.]{minted} 
\newmintinline[jl]{julia}{}
\newminted[jlcode]{julia}{linenos, firstnumber=last, breaklines, xleftmargin=\parindent}

\usepackage{subfigure}

\usepackage[english]{babel}
\usepackage{textgreek}
\usepackage{graphicx}

\usepackage{hyperref}

\usepackage[dvipsnames]{xcolor}
\usepackage[normalem]{ulem} 
\usepackage{etoolbox} 
\newbool{comments}

 \booltrue{comments} 
\ifbool{comments}{  
    
    \newcommand{\acomment}[1]{\textcolor{ForestGreen}{\emph{#1}}}

    \newcommand{\mcomment}[1]{\textcolor{NavyBlue}{\emph{#1}}}
    
    \newcommand{\lcomment}[1]{\textcolor{RedOrange}{\emph{#1}}}
    
}{

    \newcommand{\acomment}[1]{}
    \newcommand{\mcomment}[1]{}
    \newcommand{\lcomment}[1]{}
    
}

\begin{document}


\title{NetworkDynamics.jl - Composing and simulating complex networks in Julia} 



\author{Michael Lindner}
\email[]{michaellindner@pik-potsdam.de}
\affiliation{Potsdam Institute for Climate Impact Research, Telegrafenberg A31, 14473 Potsdam, Germany}
\affiliation{Institute of Theoretical Physics, Technische Universit\"at Berlin, Hardenbergstr. 36, D-10623 Berlin, Germany}

\author{Lucas Lincoln}
\affiliation{Potsdam Institute for Climate Impact Research, Telegrafenberg A31, 14473 Potsdam, Germany}
\affiliation{Chair of Control Systems and Network Control Technology, Brandenburgische Technische Universit\"at, Siemens-Halske-Ring 14, D-03046 Cottbus, Germany}

\author{Fenja Drauschke}
\author{Julia M. Koulen}
\affiliation{Potsdam Institute for Climate Impact Research, Telegrafenberg A31, 14473 Potsdam, Germany}
\affiliation{Institute of Theoretical Physics, Technische Universit\"at Berlin, Hardenbergstr. 36, D-10623 Berlin, Germany}

\author{Hans Würfel}
\affiliation{Potsdam Institute for Climate Impact Research, Telegrafenberg A31, 14473 Potsdam, Germany}
\affiliation{Department of Physics, Humboldt-Universit\"at zu Berlin, Newtonstr. 15, 12489 Berlin, Germany}

\author{Anton Plietzsch}
\affiliation{Potsdam Institute for Climate Impact Research, Telegrafenberg A31, 14473 Potsdam, Germany}
\affiliation{Department of Physics, Humboldt-Universit\"at zu Berlin, Newtonstr. 15, 12489 Berlin, Germany}

\author{Frank Hellmann}
\email[]{hellman@pik-potsdam.de}
\affiliation{Potsdam Institute for Climate Impact Research, Telegrafenberg A31, 14473 Potsdam, Germany}


\date{\today}

\begin{abstract}
NetworkDynamics.jl is an easy-to-use and computationally efficient package for simulating heterogeneous dynamical systems on complex networks, written in  Julia, a high-level, high-performance, dynamic programming language. By combining state of the art solver algorithms from DifferentialEquations.jl with efficient data structures, NetworkDynamics.jl achieves top performance while supporting advanced features like events, algebraic constraints, time-delays, noise terms and automatic differentiation.
\end{abstract}


\maketitle 


\begin{quotation}
Simulations of complex systems on networks such as power grids, with many different types of node and line dynamics are challenging to build and to optimize for performance. Available software is either proprietary or mixes higher-level scripting languages with low-level compiled languages to provide a convenient user interface while maintaining computational efficacy. The emerging programming language Julia offers a unique environment that allows to achieve both within a single language while overcoming some of the limitations of previous approaches. In this paper we present our open-source package NetworkDynamics.jl that opens up Julia's potential to the network science community. 

\noindent This article is based on {NetworkDynamics.jl v0.5}.

\end{quotation}

\section{Introduction}

The last decades have seen an ever growing need for deeply understanding dynamical systems on complex networks, with applications ranging from power grids\cite{anvari2020introduction} to neuroscience\cite{baldi1994delays, bassett2017network} and epidemic spreading\cite{bogua2003epidemic}. These systems are characterized by local components interacting with each other depending on their connection in an associated network. Examples include consumers and producers of energy that are connected via power transmission lines, neurons interacting via synapses and people transmitting diseases to new populations by traveling from city to city.

Usually such systems are high-dimensional consisting of hundreds or thousands of components. In the simplest case the dynamical laws describing the components are identical across the network, but often that is not the case and the local systems are heterogeneous with respect to their parametrization or their governing laws. The complexity of the overall system arises from the irregular coupling structure between the subsystems. The networks that represent this structure are neither regular like a grid nor completely random. It has repeatedly been observed that small changes in the network topology may have large impacts on the dynamical properties of the coupled system~\cite{menck2014dead, schultz2014detours}.

Just as in common complex systems the dynamics on network components may be modeled by discrete maps or differential equations. In the following we will focus on models consisting of ordinary differential equations (ODEs) and variations thereof introducing additional algebraic constraints, stochasticity, time-delays or events and will use the broader term of dynamical systems on networks to refer to such ODE-based models.

Methods for analytical and conceptual insight into dynamical systems on networks are well established within the research community and include master stability functions\cite{pecora1998master, borner2020delay}, linear stability theory, basin stability~\cite{menck2013basin, lindner2019stochastic}, bifurcation analysis\cite{gelbrecht2020monte} and linear response theory \cite{zhang2019fluctuation, plietzsch2019generalized}. Many of which have been developed for general dynamical systems and then successfully extended to the case of networked systems. 

However, the high complexity of coupled dynamical systems limits the insight that can be gained from purely analytical approaches and hence high-quality numerical simulations are needed. The structure of the problem leads to several difficulties that a simulation has to deal with: coupled dynamical systems are usually defined on a high-dimensional phase space, often the asymptotic properties of the system are of interest leading to a need for long integration times, subsystems may contain algebraic constraints or exhibit chaotic dynamics, interactions may introduce a time delay or the system might be subject to noise. Usually numerical experiments have to be repeated over a wide range of initial conditions and parametrizations and hence simulations have to be fast and reliable. Most of these challenges have well established algorithmic solutions; however, these often require expert knowledge of computational methods and compiled programming languages and are usually not easily accessible to non-specialists users. On the other hand, lack of these methods, and more generally of high-performance code, imposes harsh restrictions on the size and complexity of systems that can be studied and thereby limits the scientific imagination and productivity.

 A number of packages have emerged in recent years that try to enable network scientists to study more computationally intensive systems and observables, while spending less time and effort on implementing their models from scratch or adapting poorly maintained legacy code. Their common characteristic is that they offer a simplified user interface in a scripting language (Python) while relying on compiled languages as efficient computational backends (C, C++, Fortran)~\cite{rothkegel2012conedy, ansmann2018efficiently, clewley2012hybrid}. This design typically limits the interoperability with other scientific software as coupling has to happen at the slow scripting level, as well as preventing users from flexibly extending the libraries in the language that they are working in. With similar goals of facilitating network science in mind we developed the package NetworkDynamics.jl in order to leverage the potential of the Julia\cite{bezanson2017julia} ecosystem for network science while avoiding the drawbacks of two language implementations.

Julia is a new high-level, high-performance, dynamic programming language specifically tailored towards numerical computing. It is easy to write, yet is just-in-time (JiT) compiled to efficient machine code and thereby solves the two language problem, that says one must prototype in one language and then rewrite in another language for speed or deployment~\cite{bezanson2017julia}.
With the  DifferentialEquations.jl\cite{rackauckas2017differentialequations} it includes a state-of-the-art suite of differential equation solvers offering a unified user interface for solving and analyzing differential equations while not sacrificing features or performance.  Among the types of problems that can be handled by this library are stochastic differential equations (SDEs), delay differential equations (DDEs),  differential algebraic equations (DAEs), stiff equations, DEs with events, all of those combined and many more. An extensive comparison of available software for solving differential equations in various programming languages  is given by \citet{rackauckas2020comparison}. 

While it is easy to write simple networked systems in pure Julia, optimizing the system's runtime or specifying more complicated dynamics involving heterogeneities can be tricky. Here NetworkDynamics.jl comes into play with a convenient interface for the user to define local dynamics on network components on a high level of abstraction. The package uses custom-build graph data structures tailored to dynamical problems in order to construct an optimized, allocation-free Julia function that can efficiently be solved with DifferentialEquations.jl. Since composability with other packages from the Julia ecosystem is a key design principle of NetworkDynamics.jl many exciting combinations are possible. For example Julia's graph library LightGraphs.jl\footnote{\url{https://github.com/JuliaGraphs/LightGraphs.jl}} can be used to construct the input topology and since most of the solvers from DifferentialEquations.jl are compatible with automatic differentiation efficient gradients of observables with respect to input parameters can be obtained. In combination with the DiffEqFlux.jl~\cite{rackauckas2019diffeqflux} library this enables not only the use of nonlinear optimizers on the coupled system, but also the integration  and training of artificial neural networks as replacement for unknown node or edge dynamics in spirit of universal differential equations~\cite{rackauckas2020universal}.

The unique composability of different packages in the Julia open-source ecosystem enables NetworkDynamics.jl to not only be more feature complete than solutions in other languages but also faster. In a benchmark problem involving long-run simulation of a network of Kuramoto oscillators the package is shown to outperform existing software.

This paper is structured as follow: Section~\ref{sec:structure} explains the design principles that guided the development as well as the general workflow and software structure of the package, Section~\ref{sec:Julia} introduces a paradigmatic example system of Kuramoto oscillators and shows how to solve it in pure Julia, Section~\ref{sec:getting-started} guides through an implementation of the same system using NetworkDynamics.jl and several relevant generalization of it introducing heterogeneous nodes, algebraic constraints, time delays and events. In Section~\ref{sec:comparison} existing software is reviewed and in a comparison NetworkDynamics.jl is shown to significantly improve computation time of a relevant benchmark problem.

\section{Implementation strategy and software structure\label{sec:structure}}

\begin{figure*}

  \includegraphics[width=0.82\textwidth]{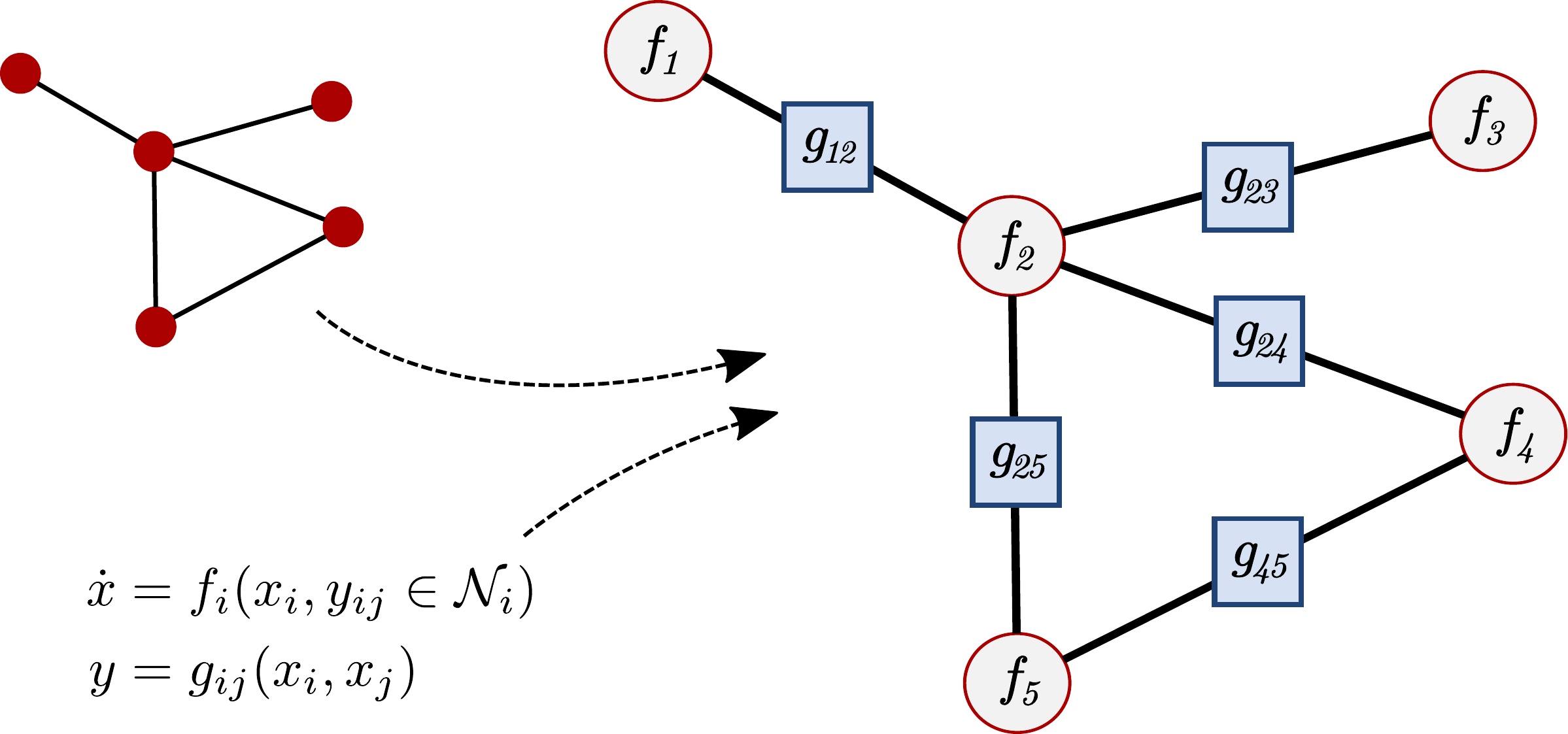}
  \caption{A diagram showing the relationship between the graph abstraction model used by NetworkDynamics.jl, a traditional mathematical graph, and the underlying equations. $\mathcal{N}_i$ denotes the set of variables of neighbors of node $i$. Nodes are connected to arbitrarily many edges, while each edge is connected to exactly two nodes. The equations are given for node dynamics described by ODEs coupled via algebraic interactions. Many modifications of this basic setup are possible. \label{fig:NDdiagram}}
\end{figure*}

NetworkDynamics.jl is designed to simulate dynamical systems consisting of coupled subsystems (sometimes referred to as \emph{components}) on a graph structure. We assume that all subsystems are either nodes (connected to an arbitrary number of edges) or edges (connected to exactly two nodes). Edges transmit interactions between their connected nodes and may be given as algebraic functions of node states or as differential equations with their own dynamic states. Understood in this sense a graph does not directly connect nodes to each other, but instead connects nodes to edges, and edges to nodes. This abstraction, which is a core concept used by NetworkDynamics.jl, provides ample flexibility to represent a wide variety of system types, while presenting a pragmatic abstraction. For a schematic view of the graph abstraction model we use compare Figure~\ref{fig:NDdiagram}.

It allows us to reduce the computation of the coupled dynamics to two core loops that iterate first over all edge components and then over all node components. If desired, these loops are almost trivially parallelizable.

The central constructor of NetworkDynamics.jl is the aptly named function \jl{network_dynamics} that receives functions for the vertex and edge subsystems as well as a graph as inputs and returns an optimized \jl{DEFunction} representing the coupled dynamical system. The graph has to be specified as a LightGraphs.jl object, a package which provides efficient implementations of general graph data structures and algorithms.

A typical workflow is as follow:

\begin{enumerate}
    \item The user specifies Julia functions for nodes and edges. These functions have to accept a fixed set of arguments corresponding to the internal state, its derivative, time and other parameters and to a set of edges for vertex functions, respectively to the source and destination vertex for edge functions. Additionally, they have to mutate their derivative and return nothing. By restricting the class of valid functions for network components, some common mistakes leading to inefficient code are incompatible by design.
    Additionally the user specifies a graph in LightGraphs.jl.
    \item NetworkDynamics.jl extracts information on the connectivity structure of the graph and pre-allocates appropriately typed caches for the vertex and edge variables. It constructs two core loops in which first the values on the edges are computed and then passed to the vertex functions. Computation of these loops is specialized towards the types of vertex and edge functions and is easily parallelizable with the corresponding keyword argument. Then the optimized \jl{DEFunction} representing the dynamical system is returned.
    \item The user specifies additional parameters for the function call and chooses an initial condition, an integration time span and an appropriate solver algorithm. The choice of solver may be automated with DifferentialEquations.jl. 
\end{enumerate}

NetworkDynamics.jl was initially developed with simulations of power systems in mind. It is the backbone of PowerDynamics.jl\cite{tkittelWIW2018,plietzsch2021powerdynamics}, an open-source framework for dynamic power grid modeling and analysis, which has been shown to be computationally more efficient than its proprietary competitors PowerFactory and MATLAB Simulink \cite{liemann2020probabilistic}. The fundamental requirement for simulating power systems is the capability of simulating large heterogeneous dynamical systems with algebraic constraints. Due to its computational efficiency and comparably simple syntax Julia was a natural choice. Additionally, DifferentialEquations.jl comes with solvers for DDEs and SDEs as well as functionalities for event handling that allow the implementation of models with delayed control schemes\cite{borner2020delay}, renewable fluctuations\cite{schmietendorf2017impact,Anvari2017} and dynamic cascading failures\cite{schafer2018dynamically}.

\subsection{Available component function types\label{sec:types}}

NetworkDynamics.jl  requires the user to specify which type of equation defines a network component. For edges those are:

\begin{itemize}
\item \jl{StaticEdge}. An edge with internal states \jl{e} that depend only on the states of the connected vertices and known parameters. 
\item \jl{StaticDelayEdge}. Like a \jl{StaticEdge} but may depend on past states of the vertices as well, used for networks  with time delays (DDEs).
\item \jl{ODEEdge}.  The evolution of the internal states \jl{e} is given by their time derivatives \jl{de}. If there are additional algebraic constraints a \jl{mass_matrix} may be specified.
\end{itemize}

And similarly for nodes:

\begin{itemize}
\item \jl{StaticVertex}. A node whose states depend only on the incoming links and known parameters. Will be transformed into an algebraic constraint in mass matrix form for the coupled differential equation.
\item \jl{ODEVertex}. The evolution of the internal states \jl{v} is specified via their time derivatives \jl{dv}. If there are additional algebraic constraints, a \jl{mass_matrix} may be specified.
\item  \jl{DDEVertex}.  Like an \jl{ODEVertex} but the derivatives may be given by delay differential equations involving past states of the internal variables.
\end{itemize}

All of these function require different calling signatures suited to the type of component they specify, e.g. a \jl{StaticVertex} is called as \jl{vertex_function!(v, edges, p, t)} with the arguments corresponding to the internal state vector, the (incoming) edges, additional parameters and time. On the other hand an  \jl{ODEVertex} requires the signature \jl{(dv, v, edges, p, t)} with the additional argument corresponding to the derivative of the internal states. For details regarding other component types have a look at our online documentation.

The central loop of the returned \jl{DEFunction} first iterates through all edge functions and then all node functions of the network. In doing so it assumes a uniform calling signature for all edge functions (and another one for all node function). As we have seen above, in heterogeneous systems the user may specify components with differing calling conventions, for example by mixing function of the types \jl{StaticVertex}, \jl{ODEVertex} and \jl{StaticDelayEdge}. In such cases NetworkDynamics.jl internally creates suitable function wrappers with a uniform calling signature. Julia's multiple-dispatch system  allows to conveniently hide such technicalities from the user under the hood of the high-level function \jl{network_dynamics} which accepts a list of \jl{VertexFunction} and a list of \jl{EdgeFunction}, regardless of their concrete type.

\subsection{Internal cache design}

To ensure efficient computations NetworkDynamics.jl pre-computes and pre-allocates as much as possible when constructing the \jl{DEFunction}. To this end several internal data structures were defined.

\jl{GraphStruct} stores all relevant static information from the network setup, including dimensions of component systems, connectivity structure, and various sets of indices into the array of dynamic variables. For every vertex these indices include information on its (incoming) edges, the dimensions of these edges, and their offsets in the internal cache array. Similar indices are stored for every edge regarding the connected nodes. \jl{GraphStruct} objects are used to initialize \jl{GraphData} objects.

\jl{GraphDataBuffer} is a mutable type storing the linearized arrays of vertex and edge variables. They get updated with the dynamic variables handled by the solver of the differential equation and with the internal variables computed by NetworkDynamics.jl.

\jl{VertexData} and \jl{EdgeData} are custom-build views that index into a subset of \jl{GraphDataBuffer}. They opt-in into Julia's fast broadcasting system and are compatible with automatic differentiation.

\jl{GraphData} stores arrays of \jl{EdgeData} and \jl{VertexData} constructed with the various sets of indies contained in \jl{GraphStuct}. This layered construction ensures that later computations can be carried out efficiently without allocating more memory on the heap, in essence providing pre-allocated views into the linearized array~\footnote{Prior to Julia v1.5 creating standard views allocated memory on the heap.}. Various accessor functions are provided, e.g. to get a list of (incoming) edges of vertex \jl{i}.

\subsection{Convenience functions\label{sec:convenience}}
Several helper function and keyword arguments are provided that allow easier setup and handling of simulations.
\begin{itemize}
\item \jl{find_fixpoint} and \jl{find_valid_ic} provide wrappers to NLsolve.jl and can be used to find fixpoints of asymptotically stable systems or to find a valid initial condition for a system with algebraic constraints (in mass matrix form).
\item \jl{network_dynamics} may be called with the keyword \jl{parallel = true} in order to enable computation of the central loop on multiple threads.
\item  \jl{coupling_sum!(dx, edges)} can be used for in-place adding the contribution of the (incoming) edges to a vertex variable.
\item When constructing a component function the user may specify symbols for the internal variables. The function \jl{syms_containing} can be used to obtain all composite symbols containing  a particular and the function \jl{idx_containing} to obtain all indices corresponding to that symbol, e.g. to extract them from the solution object for plotting. An example is shown below.
\item The package uses multiple dispatch to provide a simple interface to the user. \jl{network_dynamics} can be called either with a single vertex, respectively edge function, or with an arrays of such functions and will return \jl{DEFunction}s describing homogeneous, respectively heterogeneous systems. Depending on the data type of the graph input, this function will be optimized for directed or undirected graphs.
\item To allow for different ways to specify parameters, that \jl{ODEFunction} will behave differently depending on the type of parameters it is called with: Single numbers or Arrays will be \emph{globally} visible to every component function. When called with a tuple of parameters \jl{(vertexp, edgep)} the first element will be accessible by vertices and the second by edges only. Similarly, if \jl{vertexp} or \jl{edgep} is a number it will be passed on to all vertices or edges respectively and if it is an Array its i-th element will be passed only to the i-th vertex or edge.
\item \jl{EdgeFunction} comes with the optional keyword argument \jl{coupling} to specify if the function is directed, undirected, symmetric or antisymmetric in order to trigger suitable optimizations. A \jl{fiducial} coupling option for expert users exists as well.
\end{itemize}

\subsection{A note on edges and coupling functions}

A common way of describing a coupled dynamical system is

\begin{equation}\label{eq:ds}
     \dot x_i  = f_i(x_i) + \sum_{j=1}^N A_{ji} c_{ji}(x_j, x_i),
\end{equation}

where $x_i$ is the vector of variables at node $i$, $f_i$ is a function describing their evolution, $A$ is the adjacency matrix of the network and $c_{ji}$ describes the coupling of node $j$ to node $i$. We sometimes refer to the first argument of a coupling function as its \emph{source} and to the second as its \emph{destination}. Recall that $A_{ji}$ is $1$ if the node $j$ is connected to node $i$ and $0$ else.
Of course Equation~\eqref{eq:ds} describes only the simplest case of a coupled dynamical system and more complicated version, involving time-dependency, edge variables, noise or delay terms may be written in a similar fashion. For simplicity we will stick with this notation in this section.

Our ambition is that without compromising on performance writing code for NetworkDynamics.jl should feel to a user as if she was writing mathematical equations like Equation~\eqref{eq:ds}. This poses several challenges.

In Equation~\eqref{eq:ds} the adjacency matrix determines if an edge exists. This equation is equally valid for directed as well as for undirected graphs.  However, directed and undirected graphs differ in some intricate ways. This becomes obvious when asking, \emph{How many edges does a graph have?} 

Surprisingly, the answer to this simple question differs depending on the graph type. The number of edges $M$ of a directed graph is given by the sum of the non-zero entries of its adjacency matrix.  For an undirected graph however,  $M$ equals that sum divided by two (assuming the graph does not contain self-loops).  Indeed it makes a lot of sense to regard every edge in an undirected graph as a single edge instead of as two directed edges pointing in opposite directions. For dynamical systems on undirected networks a common and very reasonable, though rarely explicitly stated assumption is then that the coupling function is undirected as well, that is if  $A_{ji} = A_{ij}$ then $c_{ji} = c_{ij}$. Or to paraphrase: $i$ couples to $j$ through the same function as $j$ couples to $i$.

In order to simulate such a system a correspondence between edges and coupling functions has to be established. In the simplest case the coupling is \emph{homogeneous} across every edge of the network and the indices may be dropped, i.e. $c_{ij} = c$. If the coupling functions differ between edges, we found the complexity of the implementation to be lowest if for every edge exactly one coupling function is specified (which may nevertheless occur repeatedly). Even though it may seem like a technical detail, identifying $c_{ji}$ with $c_{ij}$ makes sure that such a correspondence exists  for undirected graphs.

Note however, that $c_{ji} = c_{ij}$ does not imply $c_{ji}(x_j, x_i) = c_{ji}(x_i, x_j)$, meaning the coupling function does not have to be symmetric with respect to its arguments and in most cases is not. In consequence every coupling function in an undirected graph has to be called twice, once with $x_i$ as its first and $x_j$ as its second argument, and once the other way round. Even though calling this function twice may seem to contradict our previous point about having exactly one coupling function per edge, it is in fact a consequence of the modeling approach. In a dynamical system on an undirected graph the coupling function usually models a physical conservation law holding between two nodes. While the interaction between nodes is fully described by the same law, the effect felt by each node may be different. This becomes apparent when thinking about a fluid flowing through a network of pipes and containers: The fluid flows \emph{out} of one container and \emph{into} another, affecting the respective concentrations in the containers in opposite ways. Directed graphs on the other hand, usually model asymmetric relationships, i.e. the effect of an interaction is only felt by the destination vertex. In this case the direction of the coupling function is aligned with the direction of the edge and is called only once.

NetworkDynamics.jl deals with these fundamental differences between directed and undirected graphs by dispatching on the data types \jl{SimpleDiGraph} and \jl{SimpleGraph}, provided by the graph LightGraphs.jl, and treating each case separately. This allows for the user to write her equations in an intuitive way without having to worry about order of arguments or numbers of function calls.

Internally, for directed graphs, every edge function is called exactly once with their arguments in the natural order. On the other hand, on undirected graphs a so-called \emph{fiducial orientation} is chosen for every edge, determining which of its connected vertices will be regarded as the first and which as the second argument of the coupling function. Even though we use the terms \emph{fiducial destination and source} those should not be confused with the destination and source of a flow. Those can not be known a priori since the direction of the flow is a consequence of the state of the system. The fiducial orientation on the other hand is an arbitrary choice.  In the next step, the internal dimension of every undirected edge is doubled and the coupling function is called twice, once from fiducial source to destination and once the other way round. Our data structures keep track of the fiducial orientation and resolve it by only indexing into the relevant subset of internal edge states before these states get passed on to the user-defined vertex functions. In the end, every vertex sees only its incoming edges (for directed graphs), respectively the edges for which it is the fiducial destination (for undirected graphs).

A decisive advantage of this design is that it enables optimizations for undirected networks with symmetric or anti-symmetric coupling functions. Those do not have to be called twice  since the value of the first computation can be redirected to the source as well as the destination vertex of the corresponding edge, potentially multiplied with $-1$. An important example is $c(x_j,x_i) = \sin(x_j - x_i) = - \sin(x_i - x_j) = - c(x_i,x_j)$. In order to profit from this optimization a user has to pass the keyword \jl{coupling = :antisymmetric} when constructing the \jl{EdgeFunction}.

\section{Simulating dynamics on networks in Julia\label{sec:Julia}}

An ubiquitous model of synchronization dynamics on complex network is the Kuramoto model\cite{kuramoto1975self,kuramoto2003chemical,rodrigues2016kuramoto}. It has been applied to power grids\cite{simpson2013synchronization}, neurodynamics\cite{maistrenko2007multistability}, chemical oscillators\cite{kuramoto2003chemical} and many more. The system is described by the following set of coupled ODEs.

\begin{equation}\label{eq:kuramoto}
     \dot \theta_i  = \omega_i + \sigma \sum_{j=1}^N A_{ji} \sin(\theta_j - \theta_i)
\end{equation}
Where $\theta_i$	is the phase angle of the $i$-th oscillator, $\omega_i$	is its internal frequency, $\sigma$	the coupling strength and the adjacency matrix of the network $A_{ji}$ is $1$ if the oscillators $j$ is connected to oscillator $i$ and $0$ else.

Such a system may be efficiently simulated in just a few lines of Julia code:

\begin{jlcode}
using LightGraphs, OrdinaryDiffEq

N = 10
g = watts_strogatz(N, 2, 0.)
const B = incidence_matrix(g, oriented=true)
const B_t = transpose(B)

function kuramoto_network!(dθ, θ, ω, t)
    dθ .= ω .- 5 .* (B * sin.(B_t * θ))
    return nothing
end

ω = (collect(1:N) .- sum(1:N) / N ) / N
x0 = (collect(1:N) .- sum(1:N) / N ) / N
tspan = (0., 4.)
prob = ODEProblem(kuramoto_network!, x0, tspan, ω)

sol = solve(prob, Tsit5())

\end{jlcode}

In line~1 the packages \jl{LightGraphs} for constructing a random graph and \jl{OrdinaryDiffEq} for solving the ODE are imported, then the number of vertices in the graph $N$ is defined and in line~4 a Watts-Strogatz random graph\cite{watts1998collective} is initialized. Since the average degree is chosen to be 2 and the rewiring probability 0 this amounts to a simple ring topology.   

In the function at line 8 the oriented incidence matrix \jl{B} of the graph \jl{g} is used to efficiently compute the coupling term via sparse matrix multiplication. Note however, that this simple representation of the systems dynamics is only valid if the underlying graph \jl{g} is undirected. The incidence matrix $B$ is a $N\times M$ matrix where $M$ is the number of edges and $B_{ij} = 1$ if node $i$ is the destination of edge $j$, $ 1$ if it is the source of edge $j$, and 0 else. We choose as coupling strength $\sigma = 5$. The exclamation mark \jl{!} in the function name is a Julia convention denoting functions that change ("mutate") one of their arguments and return \jl{nothing}. The dot \jl{.} is the broadcasting operator and is appended to an operator or a function in order to apply it to every element of an array.

In the following lines the internal frequencies of each oscillator are specified as a parameter array $\omega$, and then the initial condition \jl{x0} and the time span of interest is defined. Finally, everything is assembled into an \jl{ODEProblem} and solved with \jl{Tsit5} an efficient, adaptive, explicit Runge-Kutta method of order 5/4.

\begin{figure}[h]
\centering
\includegraphics[width=\columnwidth]{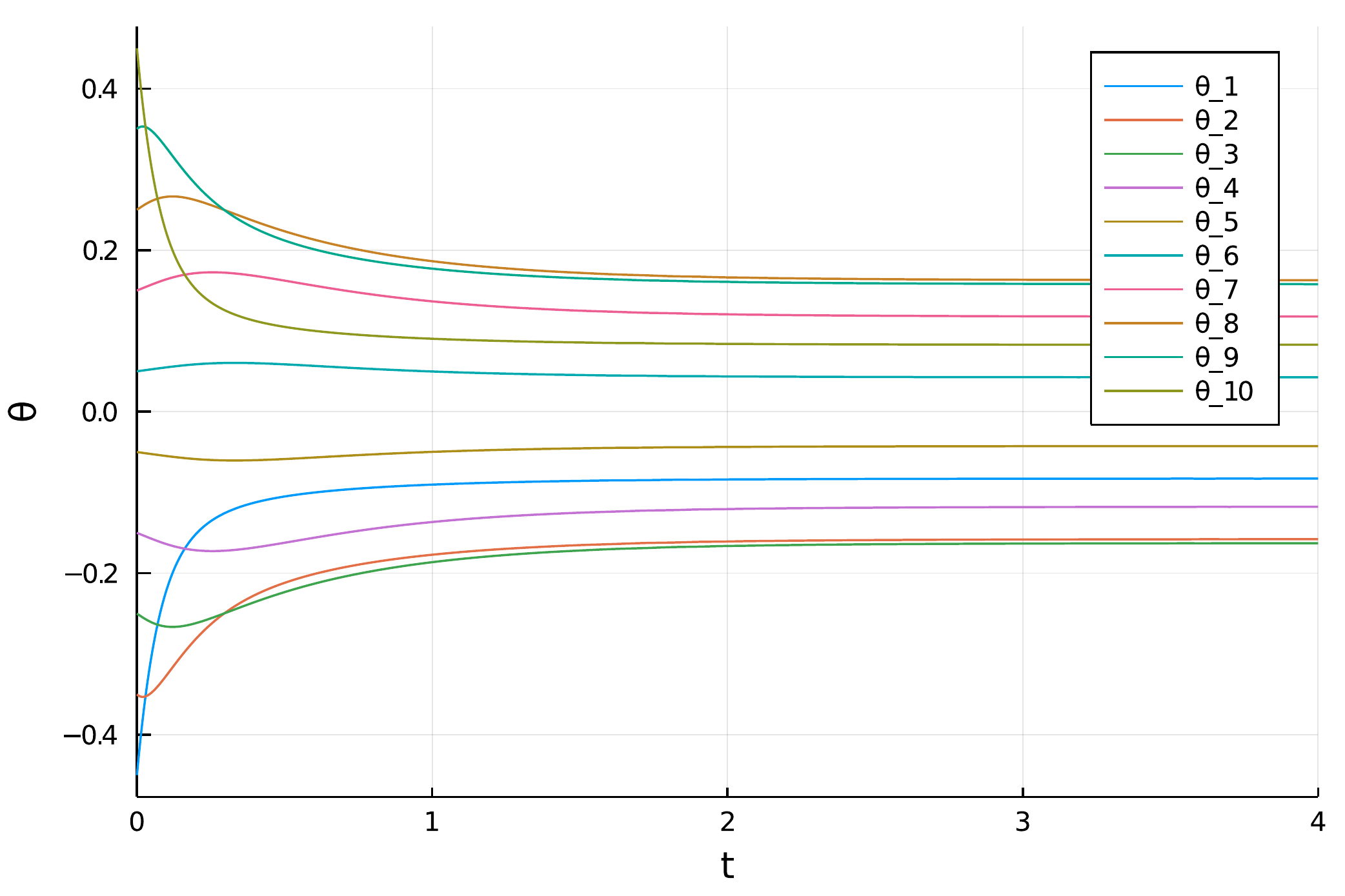}
\caption{\label{fig:homo} Phase angle dynamics of $10$ Kuramoto oscillators on a ring network with internal frequencies $\omega_i = 0.45, 0.35, 0.25, ..., -0.45$.}
\end{figure}

\section{Getting started with Network Dynamics~\label{sec:getting-started}}
The code examples shown in this section are available online\footnote{\url{https://github.com/PIK-ICoN/NetworkDynamics.jl/blob/master/examples/paper.jl}}.
In order to construct the same Kuramoto system in NetworkDynamics.jl, we start by defining generic functions for nodes and edges. By convention, these functions have to accept a specific set of arguments.

\begin{jlcode}
function kuramoto_edge!(edge, θ_s, θ_d, σ, t)
    edge .= σ .* sin.(θ_s .- θ_d)
    return nothing
end
function kuramoto_vertex!(dθ,θ,edges,ω,t)
    dθ .= ω
    for edge in edges
        dθ[1] += edge[1]
    end
    return nothing
end
\end{jlcode}

\jl{θ_s} and  \jl{θ_d} denote the internal variables of vertices which are the source and the destination of an edge respectively, while \jl{edges} denotes the arrays of (incoming) edges of a given vertex. Those arguments are always required regardless if the underlying graph is directed or undirected. For undirected graphs the notions of \emph{source} and \emph{destination} do not refer to the direction of edges in the actual graph (which are, well, undirected) but to the direction of the coupling function. 

Since generic Julia functions are allowed at network components it is necessary to supply additional information on the type of equation and the number of internal variables at a component, cf.~Section~\ref{sec:types}. To provide this information the user-specified functions are wrapped into \jl{ODEVertex} and \jl{StaticEdge} types respectively. Additionally, we specify a symbol for the vertex variables that can later be used for convenient plotting. Finally, the component functions are passed together with the graph \jl{g} to the function \jl{network_dynamics} which returns an optimized function \jl{nd!} representing the coupled ODE.

\begin{jlcode}
using NetworkDynamics

f_node! = ODEVertex(f! = kuramoto_vertex!, dim = 1, sym=[:θ])
f_edge! = StaticEdge(f! = kuramoto_edge!, dim = 1)
nd! = network_dynamics(f_node!, f_edge!, g)
\end{jlcode}

Once the \jl{ODEFunction} has been constructed everything else works as above.

\begin{jlcode}
vertexp = ω
edgep   = 5.
p = (vertexp, edgep)

nd_prob = ODEProblem(nd!, x0, tspan, p)
nd_sol = solve(nd_prob, Tsit5())
\end{jlcode}

Note that here the tuple syntax for the parameters is used, cf. Section~\ref{sec:convenience}.

The results of the simulation can be plotted straightforwardly with Plots.jl, by calling \jl{plot} on the solution object (see Fig.~\ref{fig:homo}). Note that the line labels are automatically derived from the symbol specified in the \jl{ODEVertex} type.

\begin{jlcode}
using Plots

plot(nd_sol, ylabel="θ")
\end{jlcode}

\subsection{Heterogeneous dynamics \label{sec:hetero}}

NetworkDynamics.jl was built for modeling heterogeneous system, so let us introduce some new node types now. Two paradigmatic modifications of the Kuramoto model are the Kuramoto model with inertia (also know as \emph{second order Kuramoto}) and static nodes. The second order model consists of two internal variables leading to more complicated (and for many applications more realistic) local dynamics. A static node on the other hand has no internal dynamics and instead fixes its local variable at a constant value via an algebraic equation.

\begin{jlcode}
function kuramoto_inertia!(dv,v,edges,p,t)
    dv[1] = v[2]
    dv[2] = p - v[2]
    for edge in edges
        dv[2] += edge[1]
    end
    return nothing
end

inertia! = ODEVertex(f! = kuramoto_inertia!, dim = 2, sym= [:θ, :ω])

static! = StaticVertex(
          f! = (θ, edges, c, t) -> θ .= c, 
          dim = 1, sym = [:θ])
\end{jlcode}

Since the node functions are no longer homogeneous across the network and may have a different number of local variables we have to change the coupling function slightly. Instead of using the broadcasting operator \jl{.} and applying the coupling term to every vertex variable, from here on only the first variable of each vertex system corresponding to the oscillator angle will be taken into account. Semantically this is represented by changing the names of the second and third argument to \jl{v_s, v_d} in order to denote arrays of internal vertex variables.

\begin{jlcode}
function kuramoto_edge!(edge, v_s, v_d, σ, t)
    edge[1] = σ * sin(v_s[1] - v_d[1])
    return nothing
end
\end{jlcode}

The heterogeneous  \jl{VertexFunction} have to be stored in an array of length $N$, whose indices correspond to the indices of the nodes in the network.

\begin{jlcode}
v_arr = Array{VertexFunction}( 
        [f_node! for v in vertices(g)])
v_arr[1] = inertia!
v_arr[N ÷ 2] = static!
nd_hetero! = network_dynamics(v_arr,f_edge!,g)
\end{jlcode}

Since the vertex with inertia has two internal variables we have to \jl{insert!} an initial condition for its frequency. 

\begin{jlcode}
insert!(x0, 2, 3.)
prob_hetero = ODEProblem(nd_hetero!, x0, tspan, p);
sol_hetero = solve(prob_hetero, Rodas4());
\end{jlcode}

\begin{figure}
\centering
\includegraphics[width=\columnwidth]{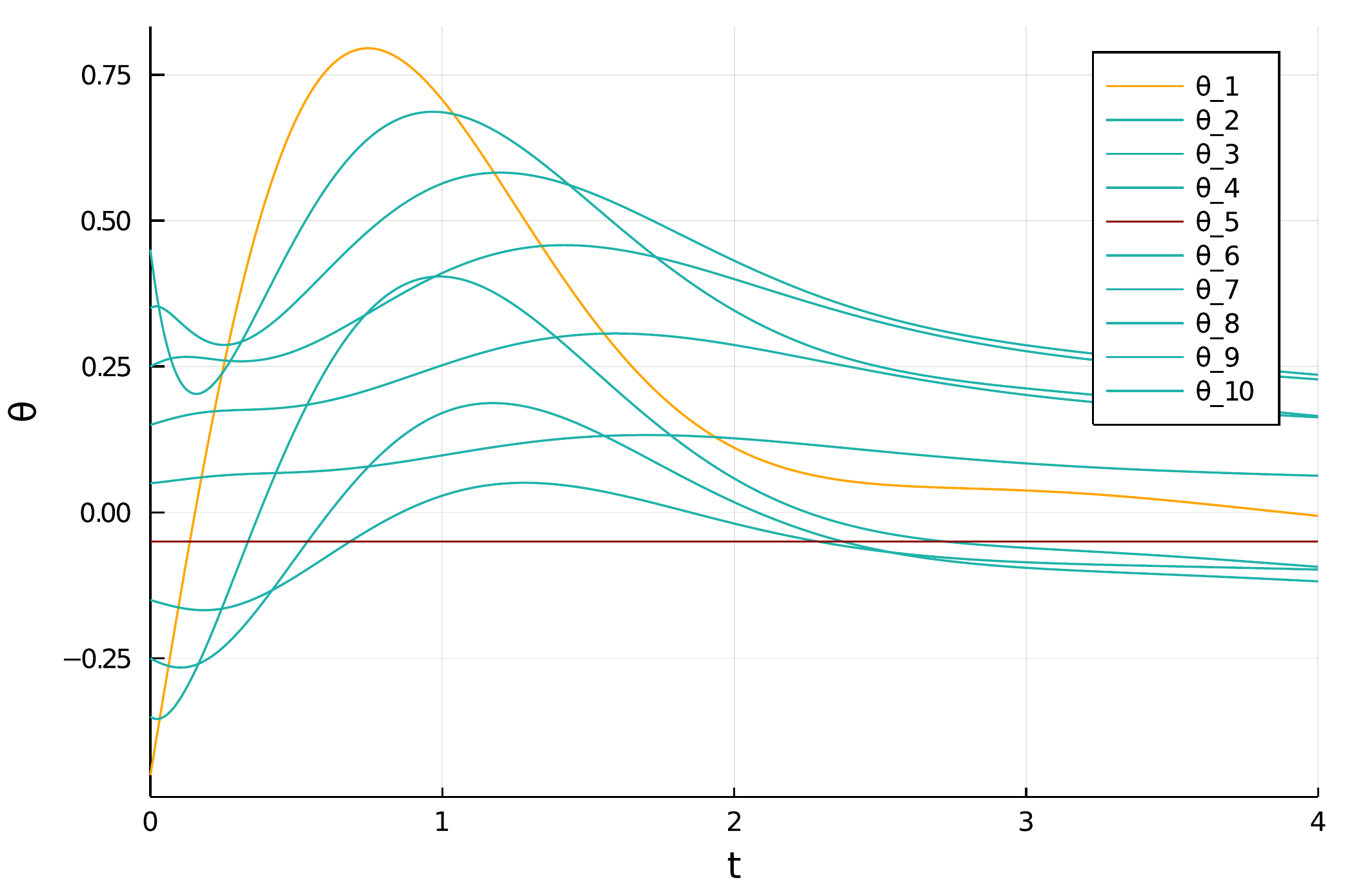}
\caption{\label{fig:hetero} Temporal evolution of a system with different node models, Node 1 is a second order Kuramoto model, Node 5 is a static node with a fixed angle $\theta$, all other nodes are the same as above, cf. Section~\ref{sec:hetero}.}
\end{figure}

Adding a static vertex that fixes the angle of one variable introduces an algebraic constraint into the ODE and we end up with a differential algebraic equation (DAE). NetworkDynamics.jl transforms the problem into a DAE in mass matrix formulation which can then be solved, e.g. by the implicit Rosenbrock method \jl{Rodas4}, and plotted just as above, see Fig.~\ref{fig:hetero}.

\subsection{Delayed coupling\label{sec:delay}}

In many real world system transmission along edges is not instantaneous but happens with a time delay $\tau$.  To this end the type \jl{StaticDelayEdge} is implemented that has access to the history \jl{h_v_s, h_v_d} of its connected vertex variables.

\begin{jlcode}
function kuramoto_delay_edge!(edge, v_s, v_d, h_v_s, h_v_d, p, t)
    edge[1] = p * sin(v_s[1] - h_v_d[1])
    return nothing
end
dedge! = StaticDelayEdge(f! = kuramoto_delay_edge!, dim = 1)

\end{jlcode}

In this case \jl{network_dynamics} returns a \jl{DDEFunction}, for which the package \jl{DelayDiffEq} provides the necessary functionality. Besides the initial condition we have to specify the delay time $\tau$ and a history function \jl{h} that returns the initial conditions for the time interval $[-\tau, 0]$. For simplicity we choose \jl{h} to be a constant extrapolation of the previous initial states \jl{x0}.

\begin{figure}
\centering
\includegraphics[width=\columnwidth]{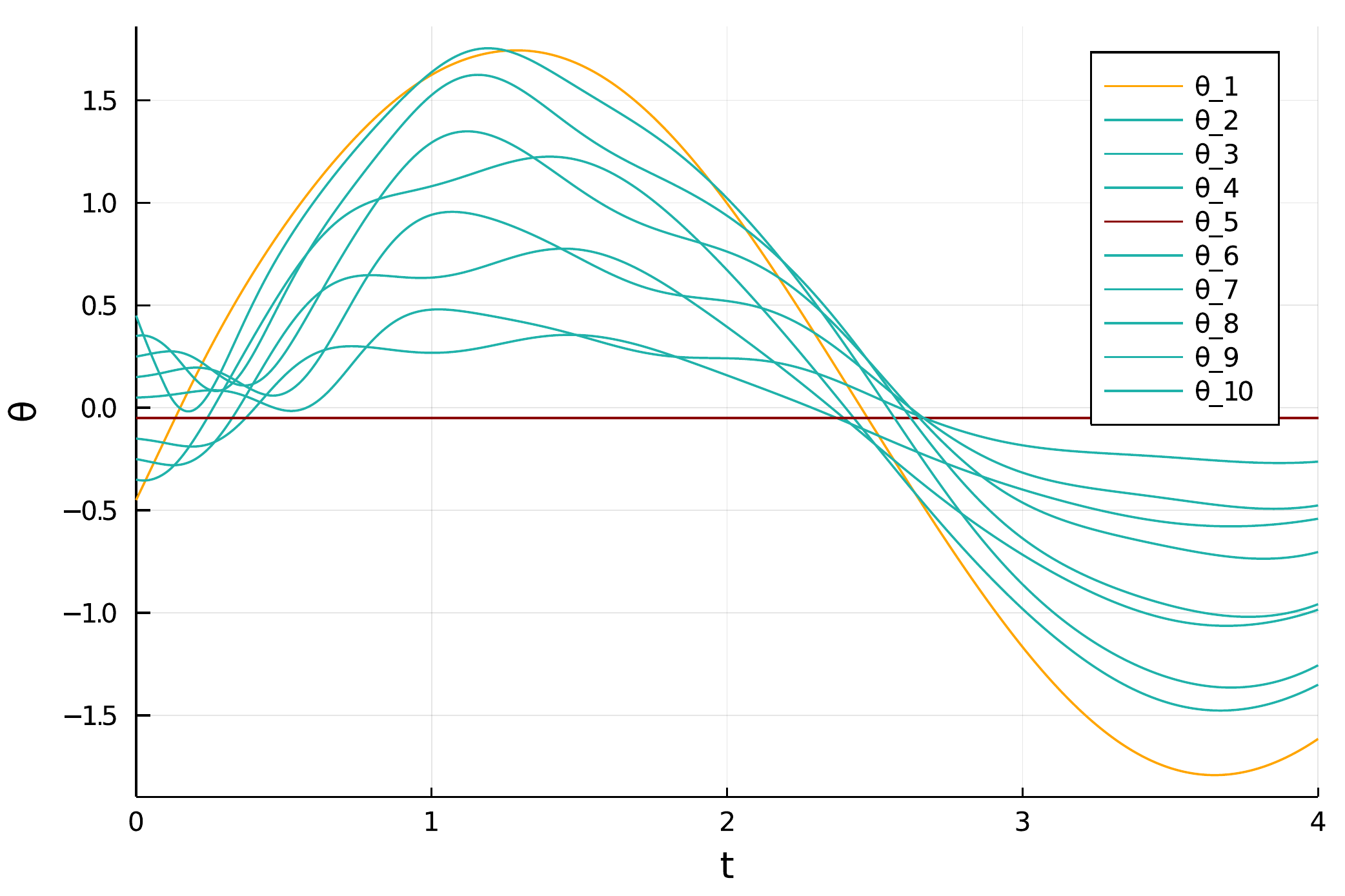}
\caption{\label{fig:delay} Same system as for Fig.~\ref{fig:hetero} but now the coupling functions have a time-delay of $\tau = 0.1$, cf. Section~\ref{sec:delay}.}
\end{figure}

\begin{jlcode}
using DelayDiffEq

h(out, p, t) = (out .= x0)
τ = 0.1
p = (vertexp,  edgep, τ)
nd_delay! = network_dynamics(v_arr,dedge!,g)
prob_delay = DDEProblem(nd_delay!, x0, h, tspan, p; constant_lags = [τ])

sol_delay = solve(prob_delay, MethodOfSteps(Rodas4(autodiff=false)))
\end{jlcode}
Here, the tuple for the parameters needs to have another field that holds the delay time. \jl{MethodOfSteps()} extends the functionality of a Julia native ODE solver to DDEs. Due to some compatibility restrictions  for Rosenbrock methods in DifferentialEquations.jl we have to specify the additional keyword \jl{autodiff = false}. The system with time delays is visualized in Fig.~\ref{fig:delay}.

\subsection{Component failures via callbacks\label{sec:callback}}
\begin{figure}
\centering
\includegraphics[width=\columnwidth]{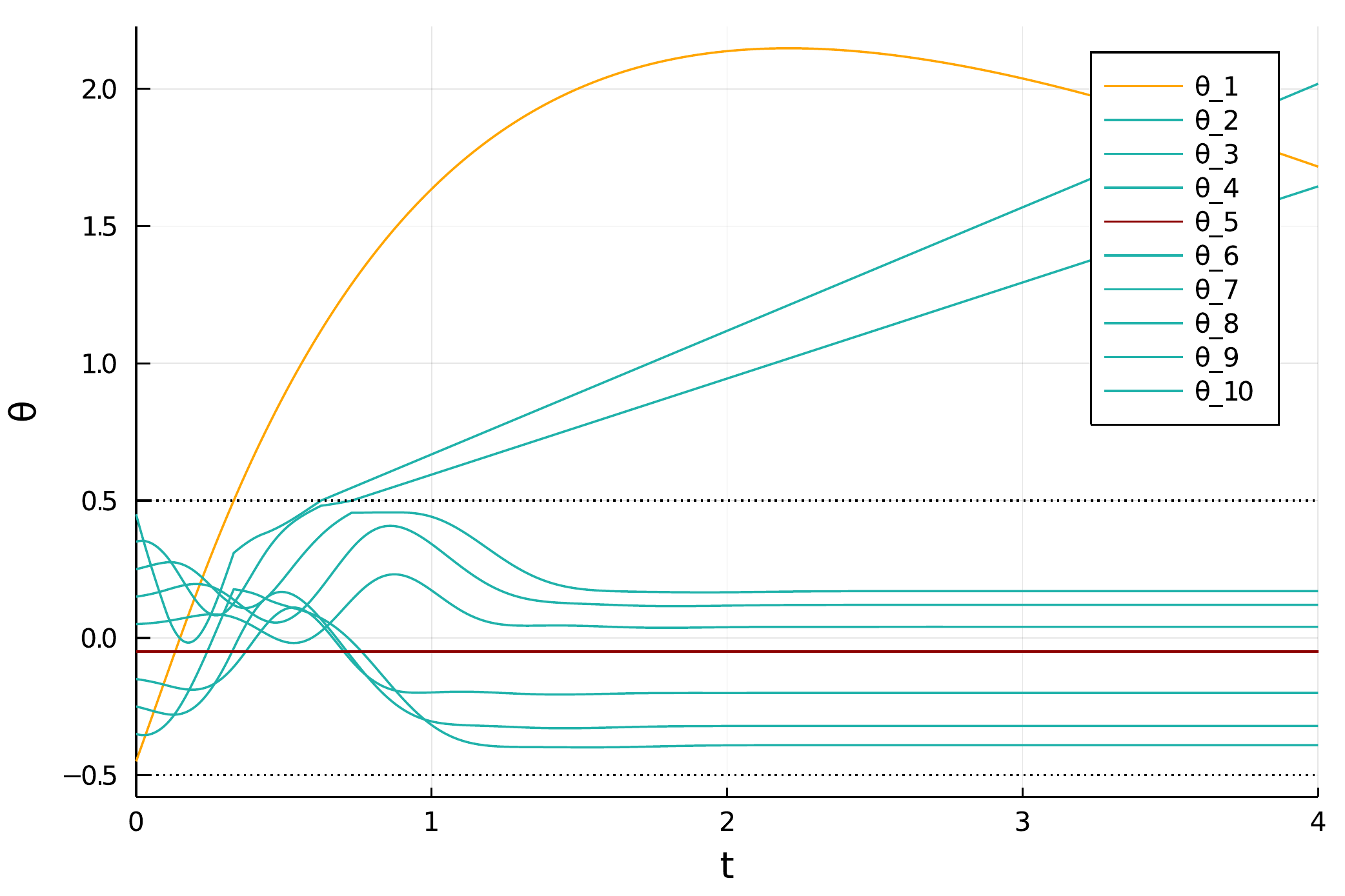}
\caption{\label{fig:callback} Same system as  as for Fig.~\ref{fig:hetero}. Here, if an oscillator leaves the interval $[-0.5, 0.5]$ it is disconnected from the rest of the system and keeps rotating according to its internal dynamics, cf. Section~\ref{sec:callback}.}
\end{figure}

The DiffEqCallbacks.jl module offers excellent support for events in various types of differential equations. As a small demonstration we model the failing of nodes in our toy model. We assume that a node fails when its internal phase leaves the "safe" interval $[-0.5,+0.5]$. When this happens all its links to neighboring nodes are cut and it keeps rotating with its internal frequency. The other nodes stay connected on a reduced grid until the next failure occurs or the system stabilizes in the "safe" interval (see Fig.~\ref{fig:callback}).

The \jl{condition} function in the next section monitors the phases $\theta$ of the different nodes. When one of its components crosses $0$, i.e. a phase leaves the "safe" interval, the \jl{affect!} is triggered. The disconnection of lines is implemented by setting their coupling strength to 0. To do this we first compute a vector whose entries are $0$ if a line is connected to the failing node, and $1$ else. This vector is in turn multiplied with the edge parameters used by the integrator when the failure occurs. Since we used a single parameter for all edges above, but now want to change the coupling strength of individual edges during the simulation we have to remake the problem with an array of edge coupling strengths.

\begin{jlcode}
using DiffEqCallbacks

θ_idxs = idx_containing(nd_delay!, :θ)

function condition(out, u, t, integrator)
  out .= (u[θ_idxs] .- 0.5) .* 
         (u[θ_idxs] .+ 0.5)
  return nothing
end

function affect!(integrator, idx)
  stable_edges = 
    map(edg -> idx ∉ edg, Pair.(edges(g)))
  integrator.p = (integrator.p[1], stable_edges .* integrator.p[2], integrator.p[3])
  return nothing
end

cb = VectorContinuousCallback(condition, affect!, 10)
prob_cb = remake(prob_delay, 
          p=(vertexp, edgep .* ones(N), τ))
sol_cb = solve(prob_cb, MethodOfSteps(Rodas4(autodiff=false)), callback=cb)

\end{jlcode}

While callbacks are fully supported more complex tasks may require detailed knowledge of our internal data structures. In order to further improve the user experience in these cases we are working on a simplified callback interface.

\subsection{Further combineable packages}

Above we saw a number of possibilities of how to make use of NetworkDynamics.jl and Julia's ecosystem for differential equations in order to solve complex tasks on complex networks. However many more is possible:
\begin{itemize}
\item StochasticDiffEq.jl~\footnote{\url{https://github.com/SciML/StochasticDiffEq.jl}} and StochasticDelayDiffEq.jl~\footnote{See \url{https://github.com/SciML/StochasticDelayDiffEq.jl} for the Julia package StochasticDelayDiffEq.jl.} provide solvers for stochastic differential equations with and without delay.
\item DynamicalSystems.jl~\cite{datseris2018dynamicalsystems} is a library targeted specifically at the complex systems community. Among others things it implements algorithms to compute Delay Embeddings and Lyapunov exponents.
\item DiffEqFlux.jl~\cite{rackauckas2019diffeqflux} combines DifferentialEquations.jl with Flux.jl, a package intended for machine learning with neural networks. Among other things this enables the use of artificial neural networks as function approximators for unknown components within coupled differential equations. Furthermore DiffEqFlux.jl provides an interface to optimization algorithms intended for high-dimensional problem such as stochastic gradient descent. An exciting application of these could be to simultaneously optimize the parameters of every subsystems with respect to a desired simulation outcome.
\item In some cases it is desirable to have a symbolic representation of a dynamical system. Making use of multiple dispatch for operators and elementary arithmetic functions ModelingToolkit.jl\cite{ma2021modelingtoolkit} implements symbolic number types that can be directly inserted into an \jl{ODEFunction} in order to obtain symbolic equations. In turn performance optimization may be applied, e.g. algebraically simplifying the equations or deriving exact Jacobians. Currently, we are working on an interface between ModelingToolkit.jl and NetworkDynamics.jl that will allow to symbolically specify network components.
\end{itemize}

\section{Software comparison\label{sec:comparison}}

As mentioned above there are a few other libraries targeted at the dynamical systems community that offer a high-level user interface while aiming for efficient computations by using code compilation. Most notably, JiTCODE~\cite{ansmann2018efficiently}  is a Python module that translates a symbolically specified user function into C Code, compiles the function, combines it with an integration algorithm and makes the compiled function available from Python as a C extension. It supports SDE, DDEs and events in combination with the packages JiTCSDE and JiTCDDE.  In a similar fashion,  PyDSTool~\cite{clewley2012hybrid} is able to compile ODEs to C Code, while Conedy~\cite{rothkegel2012conedy} is written in C++ and offers a Python interface. In a benchmark developed by \citet{ansmann2018efficiently} JiTCODE emerged as the fastest choice among these programs.

Based on their efforts we develop a paradigmatic benchmark problem in order to compare NetworkDynamics.jl to other programs solving the same system implemented in Julia,  in Python's SciPy with and without JiTCODE and in Fortran.

\subsection{Benchmark design}

Because of its widespread use in the community we choose as test system a network of Kuramoto oscillators as described by Equation~\eqref{eq:kuramoto} with random initial conditions and random eigenfrequencies of the oscillators.  The network topology is given by a Watts-Strogatz random graph~\cite{watts1998collective}, with rewiring probability $p = 0.2$, mean degree $k=4$ and varying number of nodes $N=10,\, 100,\, 1000$. These parameters lead to sparse, random, small-world networks that share some characteristics with real world-power grids~\cite{menck2013basin}.

To compare the software we compute \emph{work-precision diagrams} (WPD) for the different system sizes. WPDs are a technique from computer science, which is usually employed to benchmark different solver algorithms on the same differential equation. In our case however, both the differential equation and the solver stay the same, while implementations of them in different libraries are benchmarked against each other. The implementation strategy of the ODE for the different languages will be discussed below and as solver algorithm we use standard implementations of Dormand-Prince's 5/4 Runge-Kutta method~\cite{dormand1980family, hairer_solving_2009}. 

Depending on its parameters and initial conditions a trajectory in such a Kuramoto system may be stable, periodic or chaotic. For chaotic trajectories its hard to ensure that different implementations of the same system return the same trajectory over long integration intervals. To limit the effect of numerical errors we choose a relatively short time interval $[0,10]$ for the integration. For each system size we compute a high-precision reference trajectory in NetworkDynamics.jl  with Hairer's \jl{radau} \cite{hairer_solving_1996}  run with relative error tolerance of $10^{-10}$ and absolute error tolerance of $10^{-12}$. The error in the work precision diagrams (Fig.~\ref{fig:benchmark}-\ref{fig:benchmark3}) is reported with respect to that trajectory.

The code used for the Julia and NetworkDynamics.jl benchmarks is essentially the same as in the example outlined in Section~\ref{sec:getting-started}. For NetworkDynamics.jl we additionally used Julia's \jl{@inbounds} macro to disable bounds checking when accessing elements of arrays and explicitly declared the coupling functions as \jl{antisymmetric} which enables the reuse of coupling values where suitable. For Python we adapted the code presented in \cite{ansmann2018efficiently}. The Fortran benchmark implementation is analogous to the sparse matrix multiplication representation of the Kuramoto system used the SparseArrays.jl benchmark. All scripts used to generate the systems and plots are available online: \url{https://github.com/PIK-ICoNe/NetworkDynamicsBenchmarks}.

Execution time is measured in CPU time with Python's \jl{time} and Julia's \jl{CPUtime} modules respectively. The simulation is performed 10 times with different initial conditions and graph topologies for every system size in order to limit performance fluctuations in the simulation due to unstable trajectories and operating system states. As a result of the experiment the average simulation duration in $ms$ per oscillator is reported. All simulations were run on a Xeon CPU E5-2687W v3 @ 3.10GHz. Benchmarks used a single thread, including single threaded BLAS, for comparison purposes. Software versions are: Python 3.6.9; Julia 1.5.1; Jitcode 1.6.0; Scipy 1.5.4; NetworkDynamics 0.5.0-pre; OrdinaryDiffEq.jl 5.42.3. Jitcode uses GCC 7.5 to JiT compile, the Fortran program is compiled with gfortran 7.5. The system is running Ubuntu 18.04.

\subsection{Work-precision diagrams}

\begin{figure}
\begin{subfigure}
\centering
\includegraphics[width=\columnwidth]{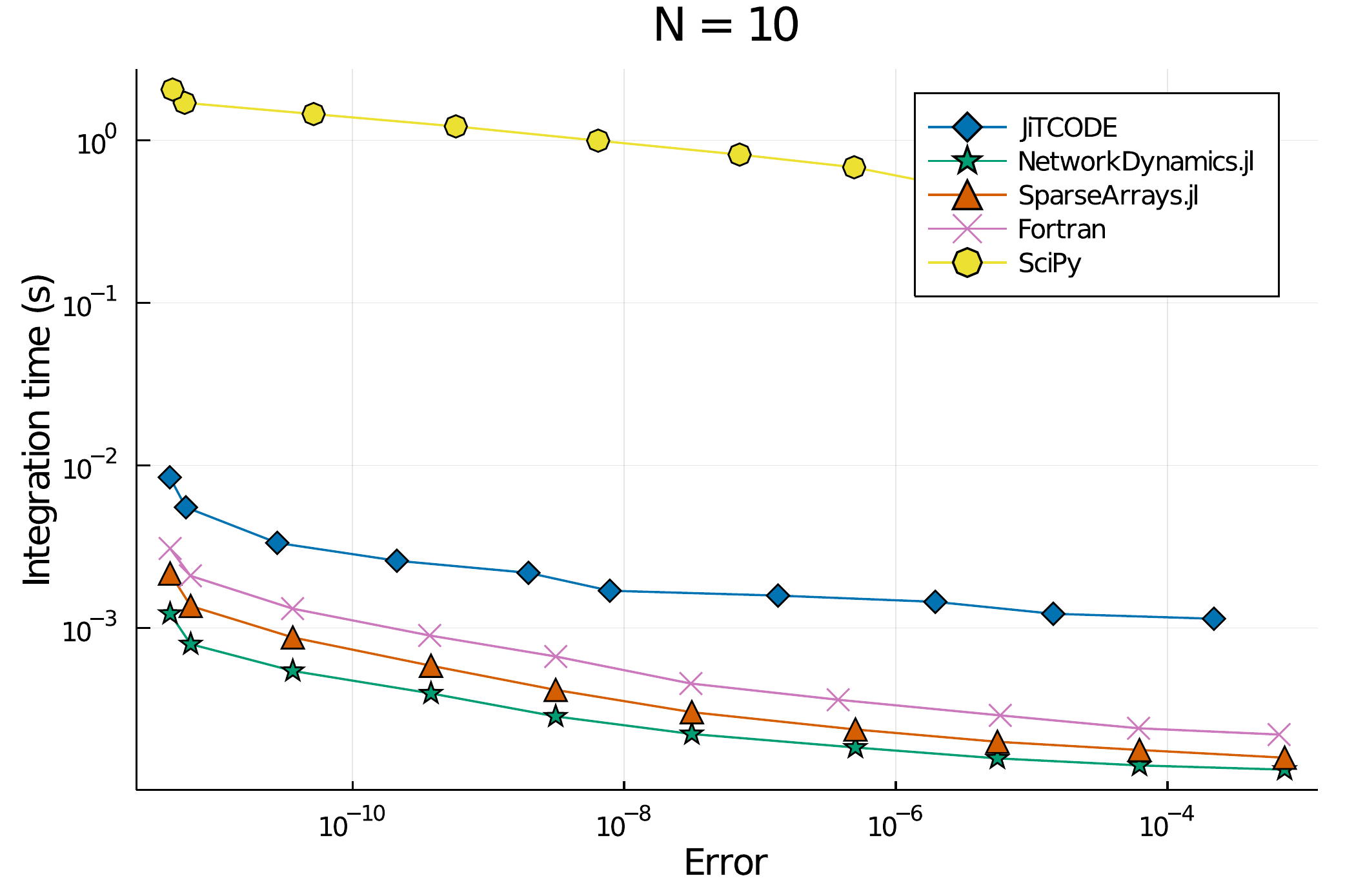}
\caption{\label{fig:benchmark} Work-precision diagram of a Kuramoto model on a Watts-Strogatz network with rewiring probability 0.2, mean degree 4 and 10 nodes. Solved with Dormand-Prince's 5th order method.}

\end{subfigure}
\begin{subfigure}
\centering

\includegraphics[width=\columnwidth]{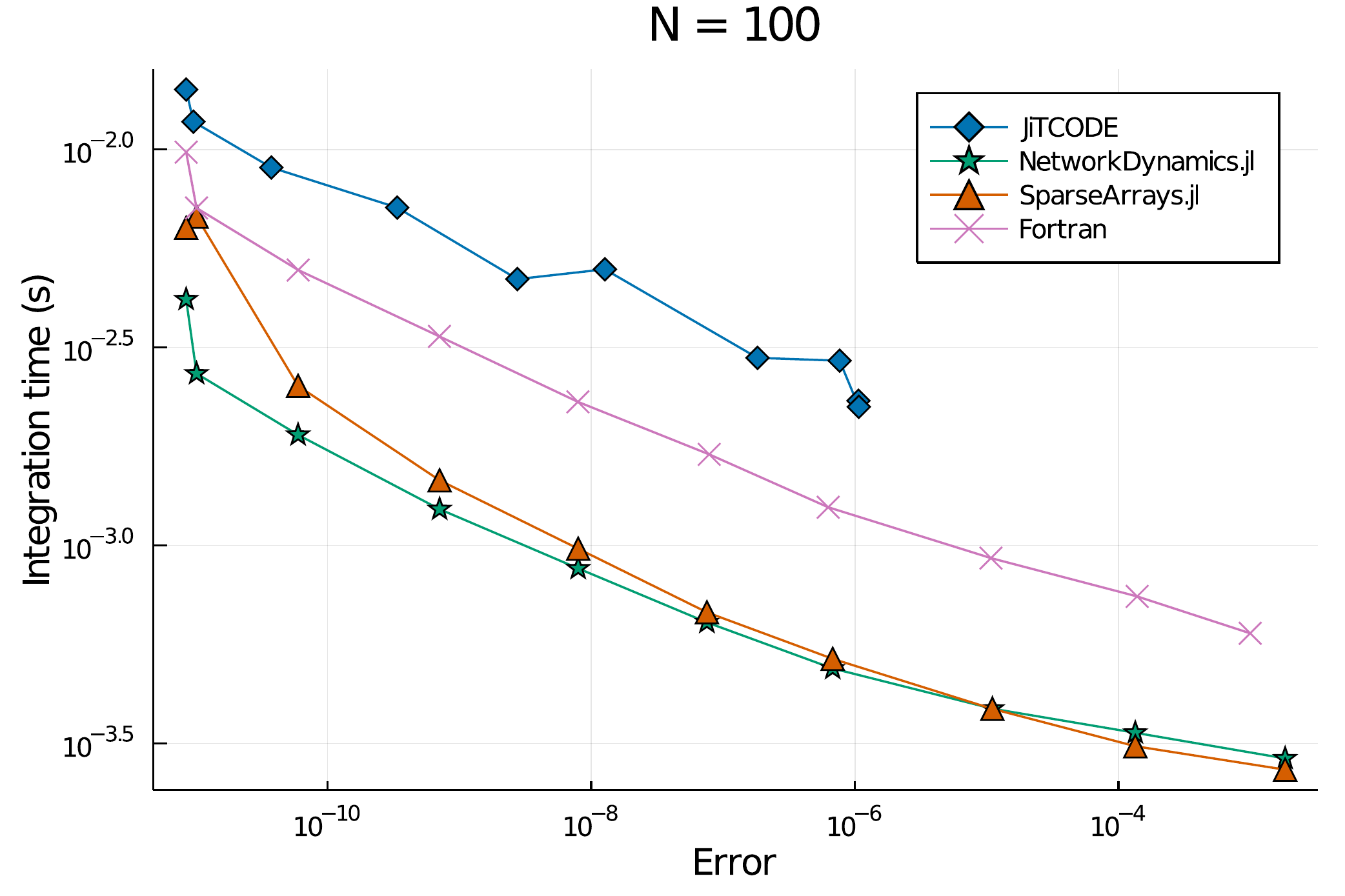}
\caption{\label{fig:benchmark2}  Same as Fig.~\ref{fig:benchmark} but with 100 nodes.}

\end{subfigure}
\begin{subfigure}
\centering

\includegraphics[width=\columnwidth]{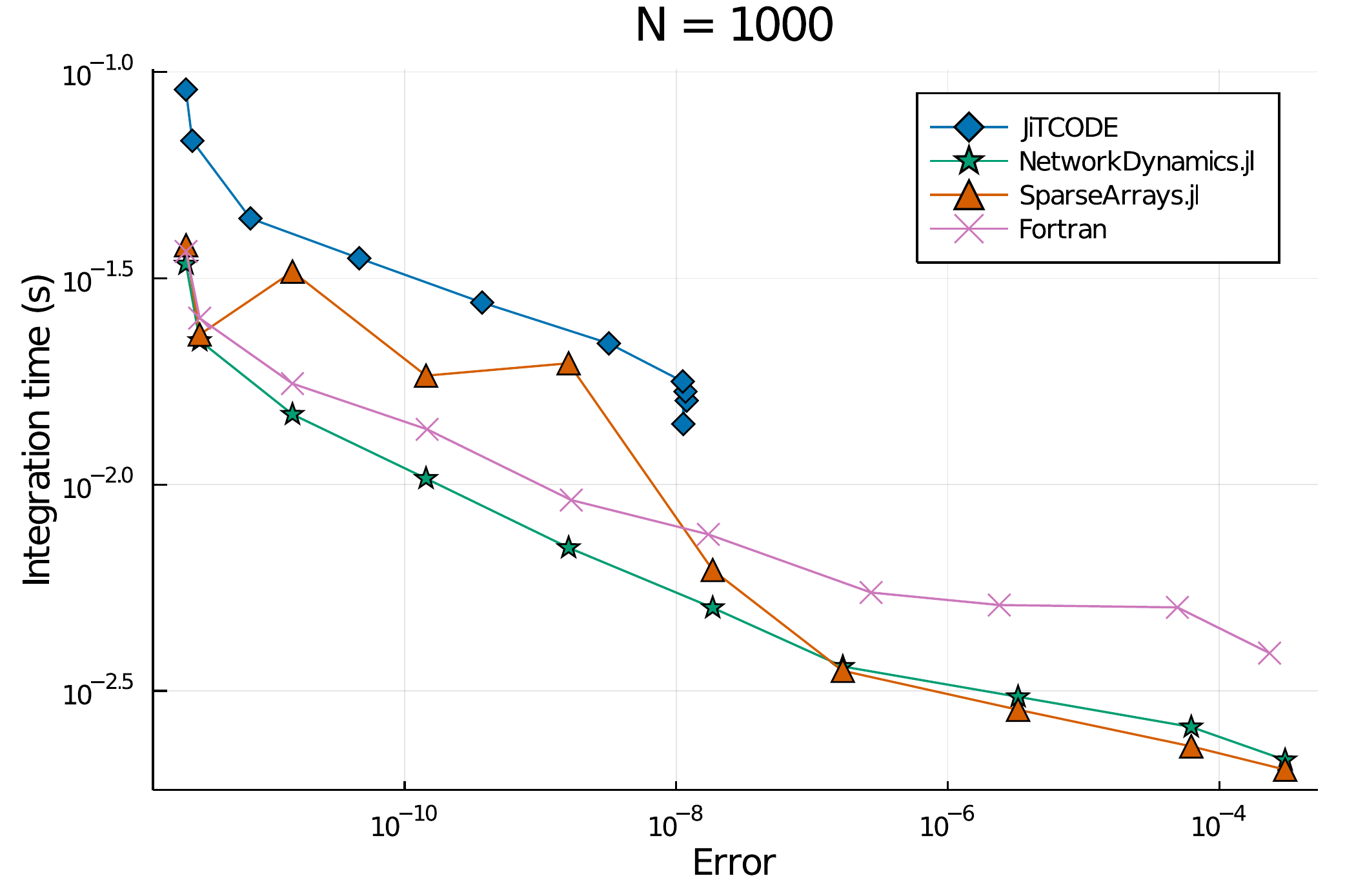}
\caption{\label{fig:benchmark3}  Same as Fig.~\ref{fig:benchmark} but with 1000 nodes. }
\end{subfigure}

\end{figure}

The work-precision diagrams show the \emph{compiled} execution time following all ahead-of-time and JiT compilation. When running a suite of experiments on the same problem set with different parameters (a common use case in our group), this is the time that each configuration following the initial takes to execute.

Both Julia implementations outperform JiTCode by factor of 2x to 4x, with a slight decrease towards larger systems. Among the reasons for the improved performs might be that the sparse data structures employed by the Julia functions make explicit use of the anti-symmetry in the coupling function. As might be expected from an interpreted language pure SciPy is considerably slower than its competitors. The performance differences range from 100x for small systems to prohibitive 10.000x for large systems. Since such a big difference distorts the scale of the plots SciPy's performance is only reported in the first WPD (Fig.~\ref{fig:benchmark}).

The performance of the Fortran version is comparable, but consistently slower, than the Julia editions, in the moderately sized systems benchmarked here. 

The experiments show that NetworkDynamics.jl is at least as fast as sparse matrix multiplication in Julia with SparseArrays.jl. This indicates that the convenient abstractions introduced by our package in order to facilitate modeling complicated networks with heterogeneous dynamics come at no additional computational cost.  Thus, users may rely on simple implementation steps as well as first-class performance even for systems for which a representation by matrix multiplication is not feasible and which would require considerable programming effort otherwise.

\subsection{Preparation time}

\begin{figure}
\centering
\includegraphics[width=\columnwidth]{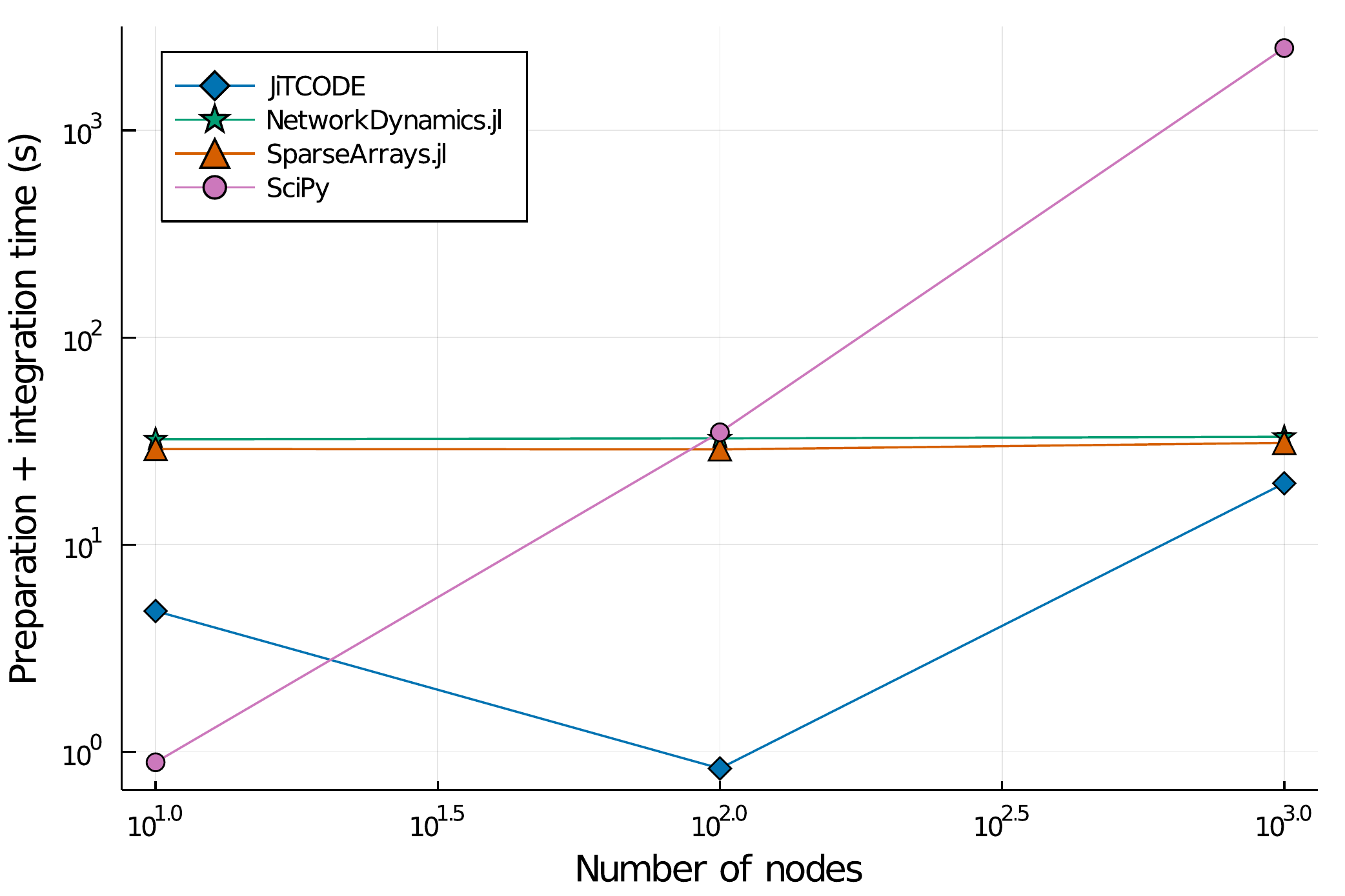}
\caption{\label{fig:sizejit} Time of the first run of each program, including composition of the network function and JiT-compilation of the solver, for different numbers of nodes in the Kuramoto network. Results are shown for relative and absolute error tolerances equal to $10^{-6}$.  }
\end{figure}

In the work-precision diagram we only reported the time it takes to run the compiled code. This ignores the time to start the runtime, import and parse all required libraries and just-in-time compile the called functions. In Fig.~\ref{fig:sizejit} we compare the time for the preparation, JiT compilation  and integration of the system for the Python and Julia based programs. Fortran performance is not shown since it is compiled ahead of time and hence the first run of the program is not expected to take longer than any other run. It is important to note that initializing the Julia runtime with the required packages \emph{once} (and performing no computations) takes nearly twice as long ($>18s$) than fully executing the compiled Fortran program (starting the process, loading experiment setup from the database, integrating the system, computing the error, and saving the results to the database) 300 times, one for each experimental condition used to generate the work-precision plots ($<10s$).

JiTCode performs this first run of the integration significantly faster than the Julia programs for small and medium sized systems. Note that the relatively long preparation time JiTCODE shows for $N=10$ nodes seems to be an artifact of  the use of SymPy over  SymEngine for small systems. This step is a lot faster for $N = 11$. For larger systems the additional cost of JiTCODE's symbolic preprocessing engine starts to dominate. Pure Scipy does not have a compilation overhead but as reported above takes a long time to integrate larger systems.

The startup time of a Julia session is by far the slowest of the programs reviewed. This is a well-known issue in the Julia community and not a specific problem of our package. More recent versions of Julia ($\geq$ v1.6) than the one available at the time of running the benchmark feature advancements in the pre-compilation stage and thus have significantly lower startup times.

We argue that, given a large number of expensive computations to be performed, the startup of the Julia runtime amortized over all trials is smaller than the speedup per trial observed compared to the Python program. This argument also holds for the Fortran version, but only in the case of a very large experimental setup, since for Fortran startup time is negligible. However, in the case of our group, the sparse-multiplication version in Fortran took more developer time than the Julia version; a trend which we suspect holds given a user equally skilled in both languages. More topically: the Fortran version does not include the abstractions available in our software NetworkDynamics.jl - a new ODE system described by a new set of equations would require only small modifications to the NetworkDynamics.jl benchmarking script, but an entire rewrite of the Fortran program.

Finally we note that it is possible to fully compile Julia programs to an executable with PackageCompiler.jl\footnote{\url{https://github.com/JuliaLang/PackageCompiler.jl}} which suffers no JiT cost, largely negating the discussion of startup time and JiT time. It is our understanding that this is not (yet) commonly done by most Julia users and as a result, we have not undertaken full binary compilation of the entire Julia benchmark program for the results in this paper.

\section{Outlook}

The Julia ecosystem has grown considerably during the last years and continues to grow at fast pace. Its focus on numerical computing and its comparably easy syntax turn it into the perfect tool for the dynamical systems community. NetworkDynamics.jl extends this functionality to dynamical simulation of heterogeneous complex networks. In this article we explained how to employ our package for highly-efficient and versatile simulations of complex networks with state-of-the-art solver algorithms. Many exciting problems that would have  required months of software research and programming just a few years ago can be constructed and solved in a few lines of code by the combination of NetworkDynamics.jl with DifferentialEquations.jl. As an example we showed how to solve a differential equation with algebraic constraints, homogeneous time-delay and state-dependent events. Many more applications that are beyond the scope of this article are possible, such as sensitive analysis via automatic differentiation or the integration of neural networks as components within coupled differential equations. Already today, NetworkDynamics.jl is the computational backend of TrussFab~\cite{kovacs2017trussfab}, a SketchUp plugin for the design of 3D truss structures and of the power systems simulator PowerDynamics.jl.  Future development goals are an interface to the symbolic modeling package ModelingToolkit.jl, support for heterogeneous time-delays and automatically deriving Jacobian-Vector product operators in order to speed up implicit solver algorithms.

\section*{Data Availability}
NetworkDynamics.jl is an open-source project and as such lives from a lively exchange of users and developers. The Julia package that supports the findings of this study is openly available on GitHub at \url{https://github.com/PIK-ICoNe/NetworkDynamics.jl}\cite{nd} and the benchmarking scripts are available at \url{https://github.com/PIK-ICoNe/NetworkDynamicsBenchmarks}\cite{ndbench}.

\begin{acknowledgments}
The authors acknowledge the support of BMBF, CoNDyNet2 FK. 03EK3055A and the Deutsche Forschungsgemeinschaft (DFG, German Research Foundation) – KU 837/39-1 / RA 516/13-1.
Mr. Lindner greatly acknowledges support by the Berlin International Graduate School in Model and Simulation based Research (BIMoS) of TU Berlin. Mr. Lincoln is supported by the German Federal Ministry for Economic Affairs and Energy (BMWi), Project No. 03EI4012A. 
All authors gratefully acknowledge the European Regional Development Fund (ERDF), the German Federal Ministry of Education and Research and the Land Brandenburg for supporting this project by providing resources on the high performance computer system at the Potsdam Institute for Climate Impact Research.
\end{acknowledgments}

\bibliography{ndjl.bib}

\providecommand{\noopsort}[1]{}\providecommand{\singleletter}[1]{#1}%
\begin{thebibliography}{47}%
\makeatletter
\providecommand \@ifxundefined [1]{%
 \@ifx{#1\undefined}
}%
\providecommand \@ifnum [1]{%
 \ifnum #1\expandafter \@firstoftwo
 \else \expandafter \@secondoftwo
 \fi
}%
\providecommand \@ifx [1]{%
 \ifx #1\expandafter \@firstoftwo
 \else \expandafter \@secondoftwo
 \fi
}%
\providecommand \natexlab [1]{#1}%
\providecommand \enquote  [1]{``#1''}%
\providecommand \bibnamefont  [1]{#1}%
\providecommand \bibfnamefont [1]{#1}%
\providecommand \citenamefont [1]{#1}%
\providecommand \href@noop [0]{\@secondoftwo}%
\providecommand \href [0]{\begingroup \@sanitize@url \@href}%
\providecommand \@href[1]{\@@startlink{#1}\@@href}%
\providecommand \@@href[1]{\endgroup#1\@@endlink}%
\providecommand \@sanitize@url [0]{\catcode `\\12\catcode `\$12\catcode
  `\&12\catcode `\#12\catcode `\^12\catcode `\_12\catcode `\%12\relax}%
\providecommand \@@startlink[1]{}%
\providecommand \@@endlink[0]{}%
\providecommand \url  [0]{\begingroup\@sanitize@url \@url }%
\providecommand \@url [1]{\endgroup\@href {#1}{\urlprefix }}%
\providecommand \urlprefix  [0]{URL }%
\providecommand \Eprint [0]{\href }%
\providecommand \doibase [0]{http://dx.doi.org/}%
\providecommand \selectlanguage [0]{\@gobble}%
\providecommand \bibinfo  [0]{\@secondoftwo}%
\providecommand \bibfield  [0]{\@secondoftwo}%
\providecommand \translation [1]{[#1]}%
\providecommand \BibitemOpen [0]{}%
\providecommand \bibitemStop [0]{}%
\providecommand \bibitemNoStop [0]{.\EOS\space}%
\providecommand \EOS [0]{\spacefactor3000\relax}%
\providecommand \BibitemShut  [1]{\csname bibitem#1\endcsname}%
\let\auto@bib@innerbib\@empty
\bibitem [{\citenamefont {Anvari}, \citenamefont {Hellmann},\ and\
  \citenamefont {Zhang}(2020)}]{anvari2020introduction}%
  \BibitemOpen
  \bibfield  {author} {\bibinfo {author} {\bibfnamefont {M.}~\bibnamefont
  {Anvari}}, \bibinfo {author} {\bibfnamefont {F.}~\bibnamefont {Hellmann}}, \
  and\ \bibinfo {author} {\bibfnamefont {X.}~\bibnamefont {Zhang}},\ }\bibfield
   {title} {\enquote {\bibinfo {title} {Introduction to focus issue: Dynamics
  of modern power grids},}\ }\href@noop {} {\bibfield  {journal} {\bibinfo
  {journal} {Chaos: An Interdisciplinary Journal of Nonlinear Science}\
  }\textbf {\bibinfo {volume} {30}},\ \bibinfo {pages} {063140} (\bibinfo
  {year} {2020})}\BibitemShut {NoStop}%
\bibitem [{\citenamefont {Baldi}\ and\ \citenamefont
  {Atiya}(1994)}]{baldi1994delays}%
  \BibitemOpen
  \bibfield  {author} {\bibinfo {author} {\bibfnamefont {P.}~\bibnamefont
  {Baldi}}\ and\ \bibinfo {author} {\bibfnamefont {A.~F.}\ \bibnamefont
  {Atiya}},\ }\bibfield  {title} {\enquote {\bibinfo {title} {How delays affect
  neural dynamics and learning},}\ }\href@noop {} {\bibfield  {journal}
  {\bibinfo  {journal} {IEEE Transactions on Neural Networks}\ }\textbf
  {\bibinfo {volume} {5}},\ \bibinfo {pages} {612--621} (\bibinfo {year}
  {1994})}\BibitemShut {NoStop}%
\bibitem [{\citenamefont {Bassett}\ and\ \citenamefont
  {Sporns}(2017)}]{bassett2017network}%
  \BibitemOpen
  \bibfield  {author} {\bibinfo {author} {\bibfnamefont {D.~S.}\ \bibnamefont
  {Bassett}}\ and\ \bibinfo {author} {\bibfnamefont {O.}~\bibnamefont
  {Sporns}},\ }\bibfield  {title} {\enquote {\bibinfo {title} {Network
  neuroscience},}\ }\href@noop {} {\bibfield  {journal} {\bibinfo  {journal}
  {Nature neuroscience}\ }\textbf {\bibinfo {volume} {20}},\ \bibinfo {pages}
  {353--364} (\bibinfo {year} {2017})}\BibitemShut {NoStop}%
\bibitem [{\citenamefont {Bogu{\'a}}, \citenamefont {Pastor-Satorras},\ and\
  \citenamefont {Vespignani}(2003)}]{bogua2003epidemic}%
  \BibitemOpen
  \bibfield  {author} {\bibinfo {author} {\bibfnamefont {M.}~\bibnamefont
  {Bogu{\'a}}}, \bibinfo {author} {\bibfnamefont {R.}~\bibnamefont
  {Pastor-Satorras}}, \ and\ \bibinfo {author} {\bibfnamefont {A.}~\bibnamefont
  {Vespignani}},\ }\bibfield  {title} {\enquote {\bibinfo {title} {Epidemic
  spreading in complex networks with degree correlations},}\ }in\ \href@noop {}
  {\emph {\bibinfo {booktitle} {Statistical mechanics of complex networks}}}\
  (\bibinfo  {publisher} {Springer},\ \bibinfo {year} {2003})\ pp.\ \bibinfo
  {pages} {127--147}\BibitemShut {NoStop}%
\bibitem [{\citenamefont {Menck}\ \emph {et~al.}(2014)\citenamefont {Menck},
  \citenamefont {Heitzig}, \citenamefont {Kurths},\ and\ \citenamefont
  {Schellnhuber}}]{menck2014dead}%
  \BibitemOpen
  \bibfield  {author} {\bibinfo {author} {\bibfnamefont {P.~J.}\ \bibnamefont
  {Menck}}, \bibinfo {author} {\bibfnamefont {J.}~\bibnamefont {Heitzig}},
  \bibinfo {author} {\bibfnamefont {J.}~\bibnamefont {Kurths}}, \ and\ \bibinfo
  {author} {\bibfnamefont {H.~J.}\ \bibnamefont {Schellnhuber}},\ }\bibfield
  {title} {\enquote {\bibinfo {title} {How dead ends undermine power grid
  stability},}\ }\href@noop {} {\bibfield  {journal} {\bibinfo  {journal}
  {Nature communications}\ }\textbf {\bibinfo {volume} {5}},\ \bibinfo {pages}
  {1--8} (\bibinfo {year} {2014})}\BibitemShut {NoStop}%
\bibitem [{\citenamefont {Schultz}, \citenamefont {Heitzig},\ and\
  \citenamefont {Kurths}(2014)}]{schultz2014detours}%
  \BibitemOpen
  \bibfield  {author} {\bibinfo {author} {\bibfnamefont {P.}~\bibnamefont
  {Schultz}}, \bibinfo {author} {\bibfnamefont {J.}~\bibnamefont {Heitzig}}, \
  and\ \bibinfo {author} {\bibfnamefont {J.}~\bibnamefont {Kurths}},\
  }\bibfield  {title} {\enquote {\bibinfo {title} {Detours around basin
  stability in power networks},}\ }\href {\doibase
  10.1088/1367-2630/16/12/125001} {\bibfield  {journal} {\bibinfo  {journal}
  {New Journal of Physics}\ }\textbf {\bibinfo {volume} {16}} (\bibinfo {year}
  {2014}),\ 10.1088/1367-2630/16/12/125001}\BibitemShut {NoStop}%
\bibitem [{\citenamefont {Pecora}\ and\ \citenamefont
  {Carroll}(1998)}]{pecora1998master}%
  \BibitemOpen
  \bibfield  {author} {\bibinfo {author} {\bibfnamefont {L.~M.}\ \bibnamefont
  {Pecora}}\ and\ \bibinfo {author} {\bibfnamefont {T.~L.}\ \bibnamefont
  {Carroll}},\ }\bibfield  {title} {\enquote {\bibinfo {title} {Master
  stability functions for synchronized coupled systems},}\ }\href@noop {}
  {\bibfield  {journal} {\bibinfo  {journal} {Physical review letters}\
  }\textbf {\bibinfo {volume} {80}},\ \bibinfo {pages} {2109} (\bibinfo {year}
  {1998})}\BibitemShut {NoStop}%
\bibitem [{\citenamefont {B{\"o}rner}\ \emph {et~al.}(2020)\citenamefont
  {B{\"o}rner}, \citenamefont {Schultz}, \citenamefont {{\"U}nzelmann},
  \citenamefont {Wang}, \citenamefont {Hellmann},\ and\ \citenamefont
  {Kurths}}]{borner2020delay}%
  \BibitemOpen
  \bibfield  {author} {\bibinfo {author} {\bibfnamefont {R.}~\bibnamefont
  {B{\"o}rner}}, \bibinfo {author} {\bibfnamefont {P.}~\bibnamefont {Schultz}},
  \bibinfo {author} {\bibfnamefont {B.}~\bibnamefont {{\"U}nzelmann}}, \bibinfo
  {author} {\bibfnamefont {D.}~\bibnamefont {Wang}}, \bibinfo {author}
  {\bibfnamefont {F.}~\bibnamefont {Hellmann}}, \ and\ \bibinfo {author}
  {\bibfnamefont {J.}~\bibnamefont {Kurths}},\ }\bibfield  {title} {\enquote
  {\bibinfo {title} {Delay master stability of inertial oscillator networks},}\
  }\href@noop {} {\bibfield  {journal} {\bibinfo  {journal} {Physical Review
  Research}\ }\textbf {\bibinfo {volume} {2}},\ \bibinfo {pages} {023409}
  (\bibinfo {year} {2020})}\BibitemShut {NoStop}%
\bibitem [{\citenamefont {Menck}\ \emph {et~al.}(2013)\citenamefont {Menck},
  \citenamefont {Heitzig}, \citenamefont {Marwan},\ and\ \citenamefont
  {Kurths}}]{menck2013basin}%
  \BibitemOpen
  \bibfield  {author} {\bibinfo {author} {\bibfnamefont {P.~J.}\ \bibnamefont
  {Menck}}, \bibinfo {author} {\bibfnamefont {J.}~\bibnamefont {Heitzig}},
  \bibinfo {author} {\bibfnamefont {N.}~\bibnamefont {Marwan}}, \ and\ \bibinfo
  {author} {\bibfnamefont {J.}~\bibnamefont {Kurths}},\ }\bibfield  {title}
  {\enquote {\bibinfo {title} {How basin stability complements the
  linear-stability paradigm},}\ }\href@noop {} {\bibfield  {journal} {\bibinfo
  {journal} {Nature physics}\ }\textbf {\bibinfo {volume} {9}},\ \bibinfo
  {pages} {89--92} (\bibinfo {year} {2013})}\BibitemShut {NoStop}%
\bibitem [{\citenamefont {Lindner}\ and\ \citenamefont
  {Hellmann}(2019)}]{lindner2019stochastic}%
  \BibitemOpen
  \bibfield  {author} {\bibinfo {author} {\bibfnamefont {M.}~\bibnamefont
  {Lindner}}\ and\ \bibinfo {author} {\bibfnamefont {F.}~\bibnamefont
  {Hellmann}},\ }\bibfield  {title} {\enquote {\bibinfo {title} {Stochastic
  basins of attraction and generalized committor functions},}\ }\href@noop {}
  {\bibfield  {journal} {\bibinfo  {journal} {Physical Review E}\ }\textbf
  {\bibinfo {volume} {100}},\ \bibinfo {pages} {022124} (\bibinfo {year}
  {2019})}\BibitemShut {NoStop}%
\bibitem [{\citenamefont {Gelbrecht}, \citenamefont {Kurths},\ and\
  \citenamefont {Hellmann}(2020)}]{gelbrecht2020monte}%
  \BibitemOpen
  \bibfield  {author} {\bibinfo {author} {\bibfnamefont {M.}~\bibnamefont
  {Gelbrecht}}, \bibinfo {author} {\bibfnamefont {J.}~\bibnamefont {Kurths}}, \
  and\ \bibinfo {author} {\bibfnamefont {F.}~\bibnamefont {Hellmann}},\
  }\bibfield  {title} {\enquote {\bibinfo {title} {Monte carlo basin
  bifurcation analysis},}\ }\href@noop {} {\bibfield  {journal} {\bibinfo
  {journal} {New Journal of Physics}\ }\textbf {\bibinfo {volume} {22}},\
  \bibinfo {pages} {033032} (\bibinfo {year} {2020})}\BibitemShut {NoStop}%
\bibitem [{\citenamefont {Zhang}\ \emph {et~al.}(2019)\citenamefont {Zhang},
  \citenamefont {Hallerberg}, \citenamefont {Matthiae}, \citenamefont
  {Witthaut},\ and\ \citenamefont {Timme}}]{zhang2019fluctuation}%
  \BibitemOpen
  \bibfield  {author} {\bibinfo {author} {\bibfnamefont {X.}~\bibnamefont
  {Zhang}}, \bibinfo {author} {\bibfnamefont {S.}~\bibnamefont {Hallerberg}},
  \bibinfo {author} {\bibfnamefont {M.}~\bibnamefont {Matthiae}}, \bibinfo
  {author} {\bibfnamefont {D.}~\bibnamefont {Witthaut}}, \ and\ \bibinfo
  {author} {\bibfnamefont {M.}~\bibnamefont {Timme}},\ }\bibfield  {title}
  {\enquote {\bibinfo {title} {Fluctuation-induced distributed resonances in
  oscillatory networks},}\ }\href@noop {} {\bibfield  {journal} {\bibinfo
  {journal} {Science Advances}\ }\textbf {\bibinfo {volume} {5}},\ \bibinfo
  {pages} {eaav1027} (\bibinfo {year} {2019})}\BibitemShut {NoStop}%
\bibitem [{\citenamefont {Plietzsch}\ \emph {et~al.}(2019)\citenamefont
  {Plietzsch}, \citenamefont {Auer}, \citenamefont {Kurths},\ and\
  \citenamefont {Hellmann}}]{plietzsch2019generalized}%
  \BibitemOpen
  \bibfield  {author} {\bibinfo {author} {\bibfnamefont {A.}~\bibnamefont
  {Plietzsch}}, \bibinfo {author} {\bibfnamefont {S.}~\bibnamefont {Auer}},
  \bibinfo {author} {\bibfnamefont {J.}~\bibnamefont {Kurths}}, \ and\ \bibinfo
  {author} {\bibfnamefont {F.}~\bibnamefont {Hellmann}},\ }\bibfield  {title}
  {\enquote {\bibinfo {title} {A generalized linear response theory of complex
  networks with an application to renewable fluctuations in microgrids},}\
  }\href@noop {} {\bibfield  {journal} {\bibinfo  {journal} {arXiv preprint
  arXiv:1903.09585}\ } (\bibinfo {year} {2019})}\BibitemShut {NoStop}%
\bibitem [{\citenamefont {Rothkegel}\ and\ \citenamefont
  {Lehnertz}(2012)}]{rothkegel2012conedy}%
  \BibitemOpen
  \bibfield  {author} {\bibinfo {author} {\bibfnamefont {A.}~\bibnamefont
  {Rothkegel}}\ and\ \bibinfo {author} {\bibfnamefont {K.}~\bibnamefont
  {Lehnertz}},\ }\bibfield  {title} {\enquote {\bibinfo {title} {Conedy: A
  scientific tool to investigate complex network dynamics},}\ }\href@noop {}
  {\bibfield  {journal} {\bibinfo  {journal} {Chaos: An Interdisciplinary
  Journal of Nonlinear Science}\ }\textbf {\bibinfo {volume} {22}},\ \bibinfo
  {pages} {013125} (\bibinfo {year} {2012})}\BibitemShut {NoStop}%
\bibitem [{\citenamefont {Ansmann}(2018)}]{ansmann2018efficiently}%
  \BibitemOpen
  \bibfield  {author} {\bibinfo {author} {\bibfnamefont {G.}~\bibnamefont
  {Ansmann}},\ }\bibfield  {title} {\enquote {\bibinfo {title} {Efficiently and
  easily integrating differential equations with jitcode, jitcdde, and
  jitcsde},}\ }\href@noop {} {\bibfield  {journal} {\bibinfo  {journal} {Chaos:
  An Interdisciplinary Journal of Nonlinear Science}\ }\textbf {\bibinfo
  {volume} {28}},\ \bibinfo {pages} {043116} (\bibinfo {year}
  {2018})}\BibitemShut {NoStop}%
\bibitem [{\citenamefont {Clewley}(2012)}]{clewley2012hybrid}%
  \BibitemOpen
  \bibfield  {author} {\bibinfo {author} {\bibfnamefont {R.}~\bibnamefont
  {Clewley}},\ }\bibfield  {title} {\enquote {\bibinfo {title} {Hybrid models
  and biological model reduction with pydstool},}\ }\href@noop {} {\bibfield
  {journal} {\bibinfo  {journal} {PLoS Comput Biol}\ }\textbf {\bibinfo
  {volume} {8}},\ \bibinfo {pages} {e1002628} (\bibinfo {year}
  {2012})}\BibitemShut {NoStop}%
\bibitem [{\citenamefont {Bezanson}\ \emph {et~al.}(2017)\citenamefont
  {Bezanson}, \citenamefont {Edelman}, \citenamefont {Karpinski},\ and\
  \citenamefont {Shah}}]{bezanson2017julia}%
  \BibitemOpen
  \bibfield  {author} {\bibinfo {author} {\bibfnamefont {J.}~\bibnamefont
  {Bezanson}}, \bibinfo {author} {\bibfnamefont {A.}~\bibnamefont {Edelman}},
  \bibinfo {author} {\bibfnamefont {S.}~\bibnamefont {Karpinski}}, \ and\
  \bibinfo {author} {\bibfnamefont {V.~B.}\ \bibnamefont {Shah}},\ }\bibfield
  {title} {\enquote {\bibinfo {title} {Julia: A fresh approach to numerical
  computing},}\ }\href@noop {} {\bibfield  {journal} {\bibinfo  {journal} {SIAM
  review}\ }\textbf {\bibinfo {volume} {59}},\ \bibinfo {pages} {65--98}
  (\bibinfo {year} {2017})}\BibitemShut {NoStop}%
\bibitem [{\citenamefont {Rackauckas}\ and\ \citenamefont
  {Nie}(2017)}]{rackauckas2017differentialequations}%
  \BibitemOpen
  \bibfield  {author} {\bibinfo {author} {\bibfnamefont {C.}~\bibnamefont
  {Rackauckas}}\ and\ \bibinfo {author} {\bibfnamefont {Q.}~\bibnamefont
  {Nie}},\ }\bibfield  {title} {\enquote {\bibinfo {title}
  {Differentialequations. jl--a performant and feature-rich ecosystem for
  solving differential equations in julia},}\ }\href@noop {} {\bibfield
  {journal} {\bibinfo  {journal} {Journal of Open Research Software}\ }\textbf
  {\bibinfo {volume} {5}} (\bibinfo {year} {2017})}\BibitemShut {NoStop}%
\bibitem [{\citenamefont {Rackauckas}(2020)}]{rackauckas2020comparison}%
  \BibitemOpen
  \bibfield  {author} {\bibinfo {author} {\bibfnamefont {C.}~\bibnamefont
  {Rackauckas}},\ }\href
  {http://www.stochasticlifestyle.com/comparison-differential-equation-solver-suites-matlab-r-julia-python-c-fortran/}
  {\enquote {\bibinfo {title} {A comparison between differential equation
  solver suites in matlab, r, julia, python, c, mathematica, maple, and
  fortran},}\ } (\bibinfo {year} {2020})\BibitemShut {NoStop}%
\bibitem [{Note1()}]{Note1}%
  \BibitemOpen
  \bibinfo {note} {\protect \url
  {https://github.com/JuliaGraphs/LightGraphs.jl}}\BibitemShut {NoStop}%
\bibitem [{\citenamefont {Rackauckas}\ \emph {et~al.}(2019)\citenamefont
  {Rackauckas}, \citenamefont {Innes}, \citenamefont {Ma}, \citenamefont
  {Bettencourt}, \citenamefont {White},\ and\ \citenamefont
  {Dixit}}]{rackauckas2019diffeqflux}%
  \BibitemOpen
  \bibfield  {author} {\bibinfo {author} {\bibfnamefont {C.}~\bibnamefont
  {Rackauckas}}, \bibinfo {author} {\bibfnamefont {M.}~\bibnamefont {Innes}},
  \bibinfo {author} {\bibfnamefont {Y.}~\bibnamefont {Ma}}, \bibinfo {author}
  {\bibfnamefont {J.}~\bibnamefont {Bettencourt}}, \bibinfo {author}
  {\bibfnamefont {L.}~\bibnamefont {White}}, \ and\ \bibinfo {author}
  {\bibfnamefont {V.}~\bibnamefont {Dixit}},\ }\bibfield  {title} {\enquote
  {\bibinfo {title} {Diffeqflux. jl-a julia library for neural differential
  equations},}\ }\href@noop {} {\bibfield  {journal} {\bibinfo  {journal}
  {arXiv preprint arXiv:1902.02376}\ } (\bibinfo {year} {2019})}\BibitemShut
  {NoStop}%
\bibitem [{\citenamefont {Rackauckas}\ \emph {et~al.}(2020)\citenamefont
  {Rackauckas}, \citenamefont {Ma}, \citenamefont {Martensen}, \citenamefont
  {Warner}, \citenamefont {Zubov}, \citenamefont {Supekar}, \citenamefont
  {Skinner},\ and\ \citenamefont {Ramadhan}}]{rackauckas2020universal}%
  \BibitemOpen
  \bibfield  {author} {\bibinfo {author} {\bibfnamefont {C.}~\bibnamefont
  {Rackauckas}}, \bibinfo {author} {\bibfnamefont {Y.}~\bibnamefont {Ma}},
  \bibinfo {author} {\bibfnamefont {J.}~\bibnamefont {Martensen}}, \bibinfo
  {author} {\bibfnamefont {C.}~\bibnamefont {Warner}}, \bibinfo {author}
  {\bibfnamefont {K.}~\bibnamefont {Zubov}}, \bibinfo {author} {\bibfnamefont
  {R.}~\bibnamefont {Supekar}}, \bibinfo {author} {\bibfnamefont
  {D.}~\bibnamefont {Skinner}}, \ and\ \bibinfo {author} {\bibfnamefont
  {A.}~\bibnamefont {Ramadhan}},\ }\bibfield  {title} {\enquote {\bibinfo
  {title} {Universal differential equations for scientific machine learning},}\
  }\href@noop {} {\bibfield  {journal} {\bibinfo  {journal} {arXiv preprint
  arXiv:2001.04385}\ } (\bibinfo {year} {2020})}\BibitemShut {NoStop}%
\bibitem [{\citenamefont {Kittel}, \citenamefont {Auer},\ and\ \citenamefont
  {Horn}(2018)}]{tkittelWIW2018}%
  \BibitemOpen
  \bibfield  {author} {\bibinfo {author} {\bibfnamefont {T.}~\bibnamefont
  {Kittel}}, \bibinfo {author} {\bibfnamefont {S.}~\bibnamefont {Auer}}, \ and\
  \bibinfo {author} {\bibfnamefont {C.}~\bibnamefont {Horn}},\ }\bibfield
  {title} {\enquote {\bibinfo {title} {Sneak preview: Powerdynamics.jl -- an
  open-source library for analyzing dynamic stability in power grids with high
  shares of renewable energy},}\ }in\ \href@noop {} {\emph {\bibinfo
  {booktitle} {17th International Workshop on Large-Scale Integration of Wind
  Power into Power Systems as well as on Transmission Networks for Offshore
  Wind Plants}}}\ (\bibinfo  {publisher} {Energynautics GmbH},\ \bibinfo {year}
  {2018})\BibitemShut {NoStop}%
\bibitem [{\citenamefont {Plietzsch}\ \emph {et~al.}(2021)\citenamefont
  {Plietzsch}, \citenamefont {Kogler}, \citenamefont {Auer}, \citenamefont
  {Merino}, \citenamefont {Gil-de Muro}, \citenamefont {Li{\ss}e},
  \citenamefont {Vogel},\ and\ \citenamefont
  {Hellmann}}]{plietzsch2021powerdynamics}%
  \BibitemOpen
  \bibfield  {author} {\bibinfo {author} {\bibfnamefont {A.}~\bibnamefont
  {Plietzsch}}, \bibinfo {author} {\bibfnamefont {R.}~\bibnamefont {Kogler}},
  \bibinfo {author} {\bibfnamefont {S.}~\bibnamefont {Auer}}, \bibinfo {author}
  {\bibfnamefont {J.}~\bibnamefont {Merino}}, \bibinfo {author} {\bibfnamefont
  {A.}~\bibnamefont {Gil-de Muro}}, \bibinfo {author} {\bibfnamefont
  {J.}~\bibnamefont {Li{\ss}e}}, \bibinfo {author} {\bibfnamefont
  {C.}~\bibnamefont {Vogel}}, \ and\ \bibinfo {author} {\bibfnamefont
  {F.}~\bibnamefont {Hellmann}},\ }\bibfield  {title} {\enquote {\bibinfo
  {title} {Powerdynamics. jl--an experimentally validated open-source package
  for the dynamical analysis of power grids},}\ }\href@noop {} {\bibfield
  {journal} {\bibinfo  {journal} {arXiv preprint arXiv:2101.02103}\ } (\bibinfo
  {year} {2021})}\BibitemShut {NoStop}%
\bibitem [{\citenamefont {Liemann}\ \emph {et~al.}(2021)\citenamefont
  {Liemann}, \citenamefont {Strenge}, \citenamefont {Schultz}, \citenamefont
  {Hinners}, \citenamefont {Porst}, \citenamefont {Sarstedt},\ and\
  \citenamefont {Hellmann}}]{liemann2020probabilistic}%
  \BibitemOpen
  \bibfield  {author} {\bibinfo {author} {\bibfnamefont {S.}~\bibnamefont
  {Liemann}}, \bibinfo {author} {\bibfnamefont {L.}~\bibnamefont {Strenge}},
  \bibinfo {author} {\bibfnamefont {P.}~\bibnamefont {Schultz}}, \bibinfo
  {author} {\bibfnamefont {H.}~\bibnamefont {Hinners}}, \bibinfo {author}
  {\bibfnamefont {J.}~\bibnamefont {Porst}}, \bibinfo {author} {\bibfnamefont
  {M.}~\bibnamefont {Sarstedt}}, \ and\ \bibinfo {author} {\bibfnamefont
  {F.}~\bibnamefont {Hellmann}},\ }\bibfield  {title} {\enquote {\bibinfo
  {title} {Probabilistic stability assessment for dynamic active distribution
  grids},}\ }in\ \href@noop {} {\emph {\bibinfo {booktitle} {2021 IEEE Madrid
  PowerTech}}}\ (\bibinfo  {publisher} {IEEE},\ \bibinfo {year} {2021})\ p.\
  \bibinfo {pages} {to appear}\BibitemShut {NoStop}%
\bibitem [{\citenamefont {Schmietendorf}, \citenamefont {Peinke},\ and\
  \citenamefont {Kamps}(2017)}]{schmietendorf2017impact}%
  \BibitemOpen
  \bibfield  {author} {\bibinfo {author} {\bibfnamefont {K.}~\bibnamefont
  {Schmietendorf}}, \bibinfo {author} {\bibfnamefont {J.}~\bibnamefont
  {Peinke}}, \ and\ \bibinfo {author} {\bibfnamefont {O.}~\bibnamefont
  {Kamps}},\ }\bibfield  {title} {\enquote {\bibinfo {title} {The impact of
  turbulent renewable energy production on power grid stability and quality},}\
  }\href@noop {} {\bibfield  {journal} {\bibinfo  {journal} {The European
  Physical Journal B}\ }\textbf {\bibinfo {volume} {90}},\ \bibinfo {pages}
  {222} (\bibinfo {year} {2017})}\BibitemShut {NoStop}%
\bibitem [{\citenamefont {Anvari}\ \emph {et~al.}(2017)\citenamefont {Anvari},
  \citenamefont {Werther}, \citenamefont {Lohmann}, \citenamefont
  {W{\"a}chter}, \citenamefont {Peinke},\ and\ \citenamefont
  {Beck}}]{Anvari2017}%
  \BibitemOpen
  \bibfield  {author} {\bibinfo {author} {\bibfnamefont {M.}~\bibnamefont
  {Anvari}}, \bibinfo {author} {\bibfnamefont {B.}~\bibnamefont {Werther}},
  \bibinfo {author} {\bibfnamefont {G.}~\bibnamefont {Lohmann}}, \bibinfo
  {author} {\bibfnamefont {M.}~\bibnamefont {W{\"a}chter}}, \bibinfo {author}
  {\bibfnamefont {J.}~\bibnamefont {Peinke}}, \ and\ \bibinfo {author}
  {\bibfnamefont {H.-P.}\ \bibnamefont {Beck}},\ }\bibfield  {title} {\enquote
  {\bibinfo {title} {Suppressing power output fluctuations of photovoltaic
  power plants},}\ }\href@noop {} {\bibfield  {journal} {\bibinfo  {journal}
  {Solar Energy}\ }\textbf {\bibinfo {volume} {157}},\ \bibinfo {pages}
  {735--743} (\bibinfo {year} {2017})}\BibitemShut {NoStop}%
\bibitem [{\citenamefont {Sch{\"a}fer}\ \emph {et~al.}(2018)\citenamefont
  {Sch{\"a}fer}, \citenamefont {Witthaut}, \citenamefont {Timme},\ and\
  \citenamefont {Latora}}]{schafer2018dynamically}%
  \BibitemOpen
  \bibfield  {author} {\bibinfo {author} {\bibfnamefont {B.}~\bibnamefont
  {Sch{\"a}fer}}, \bibinfo {author} {\bibfnamefont {D.}~\bibnamefont
  {Witthaut}}, \bibinfo {author} {\bibfnamefont {M.}~\bibnamefont {Timme}}, \
  and\ \bibinfo {author} {\bibfnamefont {V.}~\bibnamefont {Latora}},\
  }\bibfield  {title} {\enquote {\bibinfo {title} {Dynamically induced
  cascading failures in power grids},}\ }\href@noop {} {\bibfield  {journal}
  {\bibinfo  {journal} {Nature communications}\ }\textbf {\bibinfo {volume}
  {9}},\ \bibinfo {pages} {1--13} (\bibinfo {year} {2018})}\BibitemShut
  {NoStop}%
\bibitem [{Note2()}]{Note2}%
  \BibitemOpen
  \bibinfo {note} {Prior to Julia v1.5 creating standard views allocated memory
  on the heap.}\BibitemShut {Stop}%
\bibitem [{\citenamefont {Kuramoto}(1975)}]{kuramoto1975self}%
  \BibitemOpen
  \bibfield  {author} {\bibinfo {author} {\bibfnamefont {Y.}~\bibnamefont
  {Kuramoto}},\ }\bibfield  {title} {\enquote {\bibinfo {title}
  {Self-entrainment of a population of coupled non-linear oscillators},}\ }in\
  \href@noop {} {\emph {\bibinfo {booktitle} {International symposium on
  mathematical problems in theoretical physics}}}\ (\bibinfo {organization}
  {Springer},\ \bibinfo {year} {1975})\ pp.\ \bibinfo {pages}
  {420--422}\BibitemShut {NoStop}%
\bibitem [{\citenamefont {Kuramoto}(2003)}]{kuramoto2003chemical}%
  \BibitemOpen
  \bibfield  {author} {\bibinfo {author} {\bibfnamefont {Y.}~\bibnamefont
  {Kuramoto}},\ }\href@noop {} {\emph {\bibinfo {title} {Chemical oscillations,
  waves, and turbulence}}}\ (\bibinfo  {publisher} {Courier Corporation},\
  \bibinfo {year} {2003})\BibitemShut {NoStop}%
\bibitem [{\citenamefont {Rodrigues}\ \emph {et~al.}(2016)\citenamefont
  {Rodrigues}, \citenamefont {Peron}, \citenamefont {Ji},\ and\ \citenamefont
  {Kurths}}]{rodrigues2016kuramoto}%
  \BibitemOpen
  \bibfield  {author} {\bibinfo {author} {\bibfnamefont {F.~A.}\ \bibnamefont
  {Rodrigues}}, \bibinfo {author} {\bibfnamefont {T.~K.~D.}\ \bibnamefont
  {Peron}}, \bibinfo {author} {\bibfnamefont {P.}~\bibnamefont {Ji}}, \ and\
  \bibinfo {author} {\bibfnamefont {J.}~\bibnamefont {Kurths}},\ }\bibfield
  {title} {\enquote {\bibinfo {title} {The kuramoto model in complex
  networks},}\ }\href@noop {} {\bibfield  {journal} {\bibinfo  {journal}
  {Physics Reports}\ }\textbf {\bibinfo {volume} {610}},\ \bibinfo {pages}
  {1--98} (\bibinfo {year} {2016})}\BibitemShut {NoStop}%
\bibitem [{\citenamefont {Simpson-Porco}, \citenamefont {D{\"o}rfler},\ and\
  \citenamefont {Bullo}(2013)}]{simpson2013synchronization}%
  \BibitemOpen
  \bibfield  {author} {\bibinfo {author} {\bibfnamefont {J.~W.}\ \bibnamefont
  {Simpson-Porco}}, \bibinfo {author} {\bibfnamefont {F.}~\bibnamefont
  {D{\"o}rfler}}, \ and\ \bibinfo {author} {\bibfnamefont {F.}~\bibnamefont
  {Bullo}},\ }\bibfield  {title} {\enquote {\bibinfo {title} {Synchronization
  and power sharing for droop-controlled inverters in islanded microgrids},}\
  }\href@noop {} {\bibfield  {journal} {\bibinfo  {journal} {Automatica}\
  }\textbf {\bibinfo {volume} {49}},\ \bibinfo {pages} {2603--2611} (\bibinfo
  {year} {2013})}\BibitemShut {NoStop}%
\bibitem [{\citenamefont {Maistrenko}\ \emph {et~al.}(2007)\citenamefont
  {Maistrenko}, \citenamefont {Lysyansky}, \citenamefont {Hauptmann},
  \citenamefont {Burylko},\ and\ \citenamefont
  {Tass}}]{maistrenko2007multistability}%
  \BibitemOpen
  \bibfield  {author} {\bibinfo {author} {\bibfnamefont {Y.~L.}\ \bibnamefont
  {Maistrenko}}, \bibinfo {author} {\bibfnamefont {B.}~\bibnamefont
  {Lysyansky}}, \bibinfo {author} {\bibfnamefont {C.}~\bibnamefont
  {Hauptmann}}, \bibinfo {author} {\bibfnamefont {O.}~\bibnamefont {Burylko}},
  \ and\ \bibinfo {author} {\bibfnamefont {P.~A.}\ \bibnamefont {Tass}},\
  }\bibfield  {title} {\enquote {\bibinfo {title} {Multistability in the
  kuramoto model with synaptic plasticity},}\ }\href@noop {} {\bibfield
  {journal} {\bibinfo  {journal} {Physical Review E}\ }\textbf {\bibinfo
  {volume} {75}},\ \bibinfo {pages} {066207} (\bibinfo {year}
  {2007})}\BibitemShut {NoStop}%
\bibitem [{\citenamefont {Watts}\ and\ \citenamefont
  {Strogatz}(1998)}]{watts1998collective}%
  \BibitemOpen
  \bibfield  {author} {\bibinfo {author} {\bibfnamefont {D.~J.}\ \bibnamefont
  {Watts}}\ and\ \bibinfo {author} {\bibfnamefont {S.~H.}\ \bibnamefont
  {Strogatz}},\ }\bibfield  {title} {\enquote {\bibinfo {title} {Collective
  dynamics of ‘small-world’networks},}\ }\href@noop {} {\bibfield
  {journal} {\bibinfo  {journal} {nature}\ }\textbf {\bibinfo {volume} {393}},\
  \bibinfo {pages} {440--442} (\bibinfo {year} {1998})}\BibitemShut {NoStop}%
\bibitem [{Note3()}]{Note3}%
  \BibitemOpen
  \bibinfo {note} {\protect \url
  {https://github.com/PIK-ICoN/NetworkDynamics.jl/blob/master/examples/paper.jl}}\BibitemShut
  {NoStop}%
\bibitem [{Note4()}]{Note4}%
  \BibitemOpen
  \bibinfo {note} {\protect \url
  {https://github.com/SciML/StochasticDiffEq.jl}}\BibitemShut {NoStop}%
\bibitem [{Note5()}]{Note5}%
  \BibitemOpen
  \bibinfo {note} {See \protect \url
  {https://github.com/SciML/StochasticDelayDiffEq.jl} for the Julia package
  StochasticDelayDiffEq.jl.}\BibitemShut {Stop}%
\bibitem [{\citenamefont {Datseris}(2018)}]{datseris2018dynamicalsystems}%
  \BibitemOpen
  \bibfield  {author} {\bibinfo {author} {\bibfnamefont {G.}~\bibnamefont
  {Datseris}},\ }\bibfield  {title} {\enquote {\bibinfo {title}
  {Dynamicalsystems. jl: A julia software library for chaos and nonlinear
  dynamics},}\ }\href@noop {} {\bibfield  {journal} {\bibinfo  {journal}
  {Journal of Open Source Software}\ }\textbf {\bibinfo {volume} {3}},\
  \bibinfo {pages} {598} (\bibinfo {year} {2018})}\BibitemShut {NoStop}%
\bibitem [{\citenamefont {Ma}\ \emph {et~al.}(2021)\citenamefont {Ma},
  \citenamefont {Gowda}, \citenamefont {Anantharaman}, \citenamefont
  {Laughman}, \citenamefont {Shah},\ and\ \citenamefont
  {Rackauckas}}]{ma2021modelingtoolkit}%
  \BibitemOpen
  \bibfield  {author} {\bibinfo {author} {\bibfnamefont {Y.}~\bibnamefont
  {Ma}}, \bibinfo {author} {\bibfnamefont {S.}~\bibnamefont {Gowda}}, \bibinfo
  {author} {\bibfnamefont {R.}~\bibnamefont {Anantharaman}}, \bibinfo {author}
  {\bibfnamefont {C.}~\bibnamefont {Laughman}}, \bibinfo {author}
  {\bibfnamefont {V.}~\bibnamefont {Shah}}, \ and\ \bibinfo {author}
  {\bibfnamefont {C.}~\bibnamefont {Rackauckas}},\ }\bibfield  {title}
  {\enquote {\bibinfo {title} {Modelingtoolkit: A composable graph
  transformation system for equation-based modeling},}\ }\href@noop {}
  {\bibfield  {journal} {\bibinfo  {journal} {arXiv preprint arXiv:2103.05244}\
  } (\bibinfo {year} {2021})}\BibitemShut {NoStop}%
\bibitem [{\citenamefont {Dormand}\ and\ \citenamefont
  {Prince}(1980)}]{dormand1980family}%
  \BibitemOpen
  \bibfield  {author} {\bibinfo {author} {\bibfnamefont {J.~R.}\ \bibnamefont
  {Dormand}}\ and\ \bibinfo {author} {\bibfnamefont {P.~J.}\ \bibnamefont
  {Prince}},\ }\bibfield  {title} {\enquote {\bibinfo {title} {A family of
  embedded runge-kutta formulae},}\ }\href@noop {} {\bibfield  {journal}
  {\bibinfo  {journal} {Journal of computational and applied mathematics}\
  }\textbf {\bibinfo {volume} {6}},\ \bibinfo {pages} {19--26} (\bibinfo {year}
  {1980})}\BibitemShut {NoStop}%
\bibitem [{\citenamefont {Hairer}, \citenamefont {Nørsett},\ and\
  \citenamefont {Wanner}(2009)}]{hairer_solving_2009}%
  \BibitemOpen
  \bibfield  {author} {\bibinfo {author} {\bibfnamefont {E.}~\bibnamefont
  {Hairer}}, \bibinfo {author} {\bibfnamefont {S.~P.}\ \bibnamefont
  {Nørsett}}, \ and\ \bibinfo {author} {\bibfnamefont {G.}~\bibnamefont
  {Wanner}},\ }\href@noop {} {{\selectlanguage {english}\emph {\bibinfo {title}
  {Solving ordinary differential equations {I}: nonstiff problems}}}},\
  \bibinfo {edition} {2nd}\ ed.,\ \bibinfo {series} {Springer series in
  computational mathematics}\ No.~\bibinfo {number} {8}\ (\bibinfo  {publisher}
  {Springer},\ \bibinfo {address} {Heidelberg ; London},\ \bibinfo {year}
  {2009})\BibitemShut {NoStop}%
\bibitem [{\citenamefont {Hairer}\ and\ \citenamefont
  {Wanner}(1996)}]{hairer_solving_1996}%
  \BibitemOpen
  \bibfield  {author} {\bibinfo {author} {\bibfnamefont {E.}~\bibnamefont
  {Hairer}}\ and\ \bibinfo {author} {\bibfnamefont {G.}~\bibnamefont
  {Wanner}},\ }\href {\doibase 10.1007/978-3-642-05221-7} {{\selectlanguage
  {english}\emph {\bibinfo {title} {Solving {Ordinary} {Differential}
  {Equations} {II}}}}},\ \bibinfo {series} {Springer {Series} in
  {Computational} {Mathematics}}, Vol.~\bibinfo {volume} {14}\ (\bibinfo
  {publisher} {Springer Berlin Heidelberg},\ \bibinfo {address} {Berlin,
  Heidelberg},\ \bibinfo {year} {1996})\BibitemShut {NoStop}%
\bibitem [{Note6()}]{Note6}%
  \BibitemOpen
  \bibinfo {note} {\protect \url
  {https://github.com/JuliaLang/PackageCompiler.jl}}\BibitemShut {NoStop}%
\bibitem [{\citenamefont {Kovacs}\ \emph {et~al.}(2017)\citenamefont {Kovacs},
  \citenamefont {Seufert}, \citenamefont {Wall}, \citenamefont {Chen},
  \citenamefont {Meinel}, \citenamefont {M{\"u}ller}, \citenamefont {You},
  \citenamefont {Brehm}, \citenamefont {Striebel}, \citenamefont {Kommana}
  \emph {et~al.}}]{kovacs2017trussfab}%
  \BibitemOpen
  \bibfield  {author} {\bibinfo {author} {\bibfnamefont {R.}~\bibnamefont
  {Kovacs}}, \bibinfo {author} {\bibfnamefont {A.}~\bibnamefont {Seufert}},
  \bibinfo {author} {\bibfnamefont {L.}~\bibnamefont {Wall}}, \bibinfo {author}
  {\bibfnamefont {H.-T.}\ \bibnamefont {Chen}}, \bibinfo {author}
  {\bibfnamefont {F.}~\bibnamefont {Meinel}}, \bibinfo {author} {\bibfnamefont
  {W.}~\bibnamefont {M{\"u}ller}}, \bibinfo {author} {\bibfnamefont
  {S.}~\bibnamefont {You}}, \bibinfo {author} {\bibfnamefont {M.}~\bibnamefont
  {Brehm}}, \bibinfo {author} {\bibfnamefont {J.}~\bibnamefont {Striebel}},
  \bibinfo {author} {\bibfnamefont {Y.}~\bibnamefont {Kommana}},  \emph
  {et~al.},\ }\bibfield  {title} {\enquote {\bibinfo {title} {Trussfab:
  Fabricating sturdy large-scale structures on desktop 3d printers},}\ }in\
  \href@noop {} {\emph {\bibinfo {booktitle} {Proceedings of the 2017 CHI
  Conference on Human Factors in Computing Systems}}}\ (\bibinfo {year}
  {2017})\ pp.\ \bibinfo {pages} {2606--2616}\BibitemShut {NoStop}%
\bibitem [{\citenamefont {Hellmann}\ and\ \citenamefont {Lindner}(2020)}]{nd}%
  \BibitemOpen
  \bibfield  {author} {\bibinfo {author} {\bibfnamefont {F.}~\bibnamefont
  {Hellmann}}\ and\ \bibinfo {author} {\bibfnamefont {M.}~\bibnamefont
  {Lindner}},\ }\href {\doibase 10.5281/zenodo.4396193} {\enquote {\bibinfo
  {title} {{NetworkDynamics.jl}},}\ } (\bibinfo {year} {2020})\BibitemShut
  {NoStop}%
\bibitem [{\citenamefont {Lincoln}\ \emph {et~al.}(2021)\citenamefont
  {Lincoln}, \citenamefont {Drauschke}, \citenamefont {Koulen},\ and\
  \citenamefont {Lindner}}]{ndbench}%
  \BibitemOpen
  \bibfield  {author} {\bibinfo {author} {\bibfnamefont {L.}~\bibnamefont
  {Lincoln}}, \bibinfo {author} {\bibfnamefont {F.}~\bibnamefont {Drauschke}},
  \bibinfo {author} {\bibfnamefont {J.~M.}\ \bibnamefont {Koulen}}, \ and\
  \bibinfo {author} {\bibfnamefont {M.}~\bibnamefont {Lindner}},\ }\href
  {https://github.com/PIK-ICoN/NetworkDynamicsBenchmarks} {\enquote {\bibinfo
  {title} {{NetworkDynamicsBenchmarks}},}\ } (\bibinfo {year}
  {2021})\BibitemShut {NoStop}%
\end{thebibliography}%

\end{document}